%% file: DKSR_arXiv_v2.tex
\documentclass[a4paper,11pt]{article}
\pdfoutput=1 

\usepackage{jheppub} 
\makeatletter
\DeclareRobustCommand*{\bfseries}{%
  \not@math@alphabet\bfseries\mathbf
  \fontseries\bfdefault\selectfont
  \boldmath
}
\makeatother

\usepackage{calc}
\usepackage{rotating}
\usepackage[english]{babel}
\usepackage[utf8]{inputenc}
\usepackage{graphicx}
\usepackage{subfig}
\usepackage{float}
\usepackage{amsmath}
\usepackage{amssymb}
\usepackage{amsthm}
\usepackage{latexsym}
\usepackage{dcolumn}
\usepackage{hyperref}
\input macros_SUS.tex

\restylefloat{figure}

\title{Flavor anomalies and dark matter in SUSY with an extra U(1)}

\author[a]{Luc Darm\'e,}
\author[a]{Kamila Kowalska,}
\author[b,a]{Leszek Roszkowski}
\author[a]{and Enrico Maria Sessolo}

\affiliation{$^a$ National Centre for Nuclear Research,\\
Ho{\.z}a 69, 00-681 Warsaw, Poland\\
$^b$ Astrocent, Nicolaus Copernicus Astronomical Center Polish Academy of Sciences,\\  Bartycka 18, 00-716 Warsaw, Poland}

\emailAdd{luc.darme@ncbj.gov.pl}
\emailAdd{kamila.kowalska@ncbj.gov.pl}
\emailAdd{leszek.roszkowski@ncbj.gov.pl}
\emailAdd{enrico.sessolo@ncbj.gov.pl}

\abstract{Motivated by the recent anomalies in $b\rightarrow s$ transitions that emerged at LHCb, 
we consider a model with an $L_\mu - L_\tau$ gauge symmetry and additional vector-like fermions.
We find that by introducing supersymmetry the model can be made consistent with the long-standing deviation in the measured value of the anomalous magnetic moment of the muon, \gmtwo, and neutralino dark matter of broad mass ranges and properties. In particular, dark matter candidates include the well-known 1 TeV higgsino, which in the MSSM is typically not compatible with solutions to the \gmtwo\ puzzle.
Moreover, its spin-independent cross section could be at the origin of the recent small excess in XENON-1T data. 
We apply to the model constraints arising from flavor precision measurements and direct searches at the Large Hadron Collider and show that they do not currently exclude the relevant parameter space regions.}

\begin{document}
\maketitle
\section{Introduction\label{sec:intro}}

In the last few years several $B$-physics experiments have reported a number of deviations from the Standard Model (SM) in flavor-changing $b\to s$ transitions, 
which might possibly suggest the presence of new physics.
Anomalies have been first observed in the angular observable $P_5'$, measured at the Large Hadron Collider's $B$-meson factory LHCb\cite{Aaij:2015oid} (also confirmed by the Belle experiment\cite{Wehle:2016yoi}), 
and in the branching ratios for decays $B_s\to\phi\mu^+\mu^-$\cite{Aaij:2015esa} and $B\to K\mu^+\mu^-$\cite{Aaij:2014pli}.
While these observables are not necessarily associated with very clean signatures of new physics, as the QCD uncertainties in their calculation tend to be sizable,  
their anomalies have been backed up by reported deviations (or, more precisely, deficits) in the QCD-clean observables \rk\ and $R_{K^{\ast}}$\cite{Aaij:2014ora,Aaij:2017vbb}, 
the ratios of semi-leptonic $B$-meson decays to a kaon and a pair of electrons or muons, which
seem to additionally 
indicate a violation of lepton-flavor universality (LFUV). 
While the limited local statistical significance of the latter anomalies, $2.6$ and $2.2\,\sigma$, respectively, all but prevents one from claiming a genuine new physics 
phenomenon,\footnote{Note, however, that global effective-field-theory fits to the set all of $B$-physics anomalies\cite{Altmannshofer:2014rta,Altmannshofer:2017yso,DAmico:2017mtc,Ciuchini:2017mik} 
possibly point to a $\gsim 4\,\sigma$ preference for new physics
over the SM explanation.} it gives nonetheless rise to interesting speculations about the possible nature of the mechanisms responsible for their emergence.

Interestingly, the observation of LFUV signals that specifically involve muons can evoke enticing parallels 
with the longstanding $\sim 3.5\,\sigma$ discrepancy from the SM value observed in the measurement of the anomalous magnetic moment of the muon, \gmtwo, 
at Brookhaven\cite{Bennett:2006fi}. In light of the wave of renewed interest in the \gmtwo\ anomaly with the onset of new experiments 
at Fermilab\cite{Grange:2015fou,Chapelain:2017syu} and, in the near future, J-PARC\cite{Ishida:2009zz,Mibe:2010zz,Iinuma:2011zz,Saito:2012zz,Otani:2015jra}, 
which are expected to improve the sensitivity of the BNL experiment by a factor of four, it becomes compelling to try and address all these lepton-related anomalies within a common framework beyond the SM (BSM). 

Viable BSM scenarios explaining the LFUV $b\rightarrow s$ anomalies are generally roughly divided into three categories.
The first two categories involve tree-level interactions characterized by the exchange of a new gauge boson ($Z'$) (see\cite{Buras:2012jb,Gauld:2013qja,Buras:2013dea,Altmannshofer:2014cfa,Crivellin:2015mga,Crivellin:2015lwa} 
for early studies), or of a leptoquark\cite{Hiller:2014yaa}. 
In the former case, new scalars and vector-like (VL) fermions may also be added to the model, if the flavor-violating couplings are to be generated dynamically\cite{King:2017anf}.
Scenarios with a new $Z'$ gauge boson have been shown to be able to additionally accommodate the \gmtwo\ anomaly,
most easily when the $Z'$ is much lighter than $\sim 1\gev$ and feebly interacting\cite{Datta:2017pfz,Sala:2017ihs}, 
but also when the new states lie at, or just above, the 
electroweak symmetry-breaking (EWSB) scale\cite{Belanger:2015nma}. 
In the third category, loop-level box diagrams built out of new Yukawa couplings of BSM scalars and VL fermions can be used to fit the experimental $b\rightarrow s$ deviations\cite{Gripaios:2015gra,Arnan:2016cpy}, 
albeit the Yukawa couplings required are $\mathcal{O}(\sim 2)$, which usually signals a breakdown of perturbativity at energy scales as low as a few hundred~TeV.

On the other hand, typical box diagrams that can be constructed out of the gauge and Yukawa couplings of SM-like size, like those of the 
Minimal Supersymmetric Standard Model (MSSM), do not give large enough contributions to explain the LFUV anomalies, 
even in the case where general flavor-violating interactions are allowed\cite{Altmannshofer:2014rta,DAmico:2017mtc}. 
It has been shown that introducing R-parity violating interactions, in particular those generated by an $LQD^c$ superpotential term, 
can enhance squark-mediated loop contributions so that the $b\to s$ anomalies are accommodated\cite{Biswas:2014gga,Das:2017kfo,Earl:2018snx}. 
However, uncomfortably large values of R-parity violating couplings are required in this case. Incidentally, we note here that 
Ref.\cite{Altmannshofer:2017poe} employs a model with R-parity violation to address
another set of flavor observables, in semileptonic $b\rightarrow c$ transitions, 
whose measurements at Babar, Belle, and LHCb\cite{Lees:2013uzd,Hirose:2016wfn,Aaij:2015yra} have also shown a significant deviation from the SM. 
Since it was pointed out (e.g.,\cite{Buttazzo:2017ixm}) that combining viable explanations of the latter with the $b\rightarrow s$ anomalies via 
new gauge bosons incurs non-trivial model-building challenges, we refrain from including the 
$b\rightarrow c$ anomalies in this work.   

When it comes to the \gmtwo\ anomaly, finally, one should note that supersymmetric (SUSY) spectra featuring sleptons, 
charginos, and neutralinos in the few hundred~GeV range are known to provide a viable solution,
as long as they happen to be much lighter than the states
charged under SU(3)\cite{Akula:2013ioa,Kowalska:2015zja}, which are strongly constrained by the 125\gev\ Higgs boson mass and direct LHC searches.

One issue of great interest in the context of constructing viable SM extensions is the nature of dark matter (DM) in the Universe. 
The presence of a DM candidate can be easily accommodated  in nearly all of the models mentioned above, as long as they feature 
a discrete symmetry stabilizing the DM candidate. The latter might emerge as a remnant $\mathbb{Z}_2$ from the breaking of the new abelian gauge group in models with a $Z'$ boson, 
or be due to R-parity in SUSY models, or other. 
However, in the most basic simplified models engineered to explain the flavor anomalies 
the DM candidate does not emerge as a natural byproduct of the theory, but it is rather added by hand.
For example, in SM extensions by a new abelian gauge symmetry, which provide one of the most natural frameworks for 
the violation of flavor universality, a viable DM particle takes generally the form of an extra inert scalar field, which does not play a part in the spontaneous 
breaking of the new gauge symmetry\cite{Sierra:2015fma,Belanger:2015nma,Ko:2017yrd,Kawamura:2017ecz,Chiang:2017zkh}, or of an extra DM fermion\cite{Kile:2014jea,Celis:2016ayl,Altmannshofer:2016jzy,Cline:2017lvv,Ko:2017quv,Baek:2017sew,Cline:2017qqu,Falkowski:2018dsl}. 
From this point of view, extensions of the SM based on SUSY offer the advantage of having the DM candidate, which is already embedded naturally in the theory, in the form of the lightest neutralino.

One might consider other reasons why it is worth exploring the possibility of a SUSY extension as a viable explanation for
the $b\rightarrow s$ anomalies, \gmtwo, and the relic DM density. Besides protecting the parameters of the scalar sector against large quantum corrections,
SUSY provides a framework for the presence of a large number of particles in the spectrum, which end up greatly facilitating an explanation of 
the \gmtwo\ anomaly with much a broader range of DM mass values than in typical models not based on supersymmetry.
In fact, in typical non-SUSY $Z'$ models one is constrained  by the measurement of $\gmtwo$ to a rough upper bound for the DM mass,
$m_{\textrm{DM}}\lesssim 250\gev$\cite{Belanger:2015nma}.
Incidentally, note that this feature has been very recently employed in\cite{Banerjee:2018eaf}, where a $U(1)$-extension of the SM with two extra singlet superfields was supersymmetrized. 
Because of the large number of neutralinos in the spectra, it was shown there that the parameter space in agreement with the \gmtwo\ anomaly is also 
consistent with a larger neutralino mass scale than in the MSSM, and with data from a set of neutrino experiments that is not explained in analogous non-SUSY scenarios. 
   
In the context of the weakly interacting massive particle (WIMP) paradigm having a heavier DM is particularly appealing. Firstly, 
light WIMPs are becoming increasingly constrained by direct DM searches, both at the LHC and, especially, 
in underground Xenon detectors. The $\tev$-scale range of WIMP mass, on the other hand, is much unscathed by the current direct bounds. 
Secondly, DM at the TeV scale has emerged after the discovery of the 125\gev\ Higgs boson as a natural favorite in many models of new physics, 
and particularly in SUSY. In fact, the mass of the Higgs boson implies 
expectations for the SUSY spectrum that naturally accommodate
TeV-scale DM candidates like the $\sim 1\tev$ higgsino (see, e.g.,\cite{Kowalska:2013hha}, 
and\cite{Kowalska:2015kaa,Roszkowski:2017nbc,Kowalska:2018toh} for recent reviews).

In this regard, it is very recent news that the XENON-1T Collaboration has recorded a slight excess of events possibly corresponding to a signal in the hundreds of~GeV or a TeV range\cite{Aprile:2018dbl}. 
Again, it is way too early to become excited, but there exists a possibility that the data point towards a DM candidate consistent with the 
TeV-scale, like the $\sim 1\tev$ higgsino.

In this study we consider a supersymmetric class of BSM scenarios with a spontaneously broken U(1)$_X$ gauge symmetry and VL fermions, which can explain the LFUV experimental anomalies. As a concrete example we identify U(1)$_X$ with the anomaly-free global symmetry of the SM, $L_\mu-L_\tau$\cite{Foot:1990mn,He:1990pn,He:1991qd} 
(the difference between the muon and tau lepton numbers), following, e.g., Ref.\cite{Altmannshofer:2014cfa} and other subsequent papers. Apart from the $B$-physics anomalies, U(1)$_{L_{\mu}-L_{\tau}}$ can accommodate several other experimentally measured phenomena, like particular textures in the neutrino mass matrices\cite{Ma:2001md} 
and the discrepancy in \gmtwo\cite{Baek:2001kca,Ma:2001md,Altmannshofer:2016brv}. 
Additionally, it guarantees that the new BSM sector remains well hidden from direct searches at proton and electron colliders, 
a welcome feature given the ever increasing lower bounds on BSM particle masses.
We show that this framework resolves the \gmtwo\ discrepancy with, among others, 
a $1\tev$ higgsino DM candidate, which is not possible in the MSSM.

The structure of the paper is as follows.
In \refsec{sec:model} we introduce the model and discuss some of its most relevant features for a subsequent analysis of the $B$-physics anomalies.  In \refsec{sec:flav} we discuss in detail BSM contributions to the observables \rk\ and \rks, $B_s$ mixing, and \gmtwo\ that can arise in our model. 
The main results of the study are presented in~\refsec{sec:bench}, which contains several benchmark points, a study of their dark matter properties and bounds and prospect from LHC searches and 
electroweak precision observables.  
We summarize our findings and conclude in \refsec{sec:sum}. We provide technical details of the supersymmetric model in Appendix~\ref{app:model}.

\section{Supersymmetric model with extra gauged U(1) \label{sec:model}}

Motivated by the fact that extensions of the SM by a new gauge boson provide arguably the most natural solution
to the LHCb flavor anomalies\cite{Altmannshofer:2014cfa}, we consider in this work an extra abelian anomaly-free gauge group, $U(1)_{L_{\mu}-L_{\tau}}$ (denoted $U(1)_X$ in the rest of the paper). Under the new gauge symmetry the SM quarks remain neutral, whereas the SM leptons acquire $L_{\mu}-L_{\tau}$ charges. 
Thus, the quantum numbers of the SM lepton Weyl spinors under the SU(3)$\times$SU(2)$_L\times$U(1)$_Y\times$U(1)$_X$ group are
\bea
l_1:(\mathbf{1},\mathbf{2},-1/2,0) & \quad & e_R:(\mathbf{1},\mathbf{1},1,0)\label{electrons}\\
l_2:(\mathbf{1},\mathbf{2},-1/2,1) & \quad & \mu_R:(\mathbf{1},\mathbf{1},1,-1)\label{muons}\\
l_3:(\mathbf{1},\mathbf{2},-1/2,-1) & \quad & \tau_R:(\mathbf{1},\mathbf{1},1,1)\label{taus}\,, 
\eea
where we use a standard convention for the electron doublet, $l_1=(\nu_e,e_L)^T$ and equivalent conventions for muons and taus apply.

The U(1)$_X$ symmetry is broken by the vacuum expectation values (vevs) of 2 new scalar fields, which are singlets under the SM group, 
\be
S_1:(\mathbf{1},\mathbf{1},0,-1) \quad S_2:(\mathbf{1},\mathbf{1},0,1)\,,
\ee
and which will be promoted to left-chiral superfields after supersymmetrization of the Lagrangian. 
Note that in what follows we always indicate with lower-case letters the SM fields and with capital ones the BSM states.

In order to generate the flavor-changing coupling  of the new gauge boson, $Z'$, to the $b_L$ and $s_L$ quarks 
one needs a VL pair of SU(2)-doublet quarks, one of which will mix with the SM left-handed quarks after the breaking of the U(1)$_X$ symmetry:
\be\label{VLquarks}
Q:(\mathbf{3},\mathbf{2},1/6,-1)=
\left( {\begin{array}{c} U_L\\
D_L
 \end{array} } \right) \quad Q':(\mathbf{\bar{3}},\mathbf{2},-1/6,1)=\left( {\begin{array}{c} D_R\\
U_R
 \end{array} } \right)\,.
\ee
We also introduce two U(1)$_X$ neutral generations of VL leptons:
\be
L:(\mathbf{1},\mathbf{2},-1/2,0)=
\left( {\begin{array}{c}
N_L\\
E_L
 \end{array} } \right) \quad L':(\mathbf{1},\mathbf{2},1/2,0)=\left( {\begin{array}{c}
E_R\\
N_R
 \end{array} } \right),
\ee
\be
E:(\mathbf{1},\mathbf{1},1,0) \quad E':(\mathbf{1},\mathbf{1},-1,0)\,.
\ee
The reason for introducing new leptons is twofold. On theoretical grounds one could
argue that VL lepton families are expected in the presence of 
TeV-scale VL quarks, if the model is to be eventually compatible with some sort of Grand Unification (GUT) mechanism (we will not, however, make any attempt at a GUT completion in this work).
Moreover, the addition of VL leptons opens up extra channels for the interaction of the muons with the photon, 
which will lead to a better fit to the \gmtwo\ anomaly
over a broad range of input parameters (see also\cite{Altmannshofer:2016oaq} for a similar setup in the non-SUSY framework).

We can now write the superpotential of our model,
\bea
W&\supset& -Y_e\,l h_d e_R-Y_d\,q h_d d_R+Y_u q h_h u_R+\mu h_u h_d\nonumber\\
 & &-\lam_{E,2} S_2 E' \mu_R-\lam_{E,3} S_1 E' \tau_R-\widetilde{Y}_1\,L\,h_d E + \widetilde{Y}_2 L' h_u E'-\widetilde{Y}_{l_1} l_1 h_d E-\widetilde{Y}_{e_1} L\,h_d e_R\nonumber\\
 & &-\lam_{L,2} S_1 l_2\,L'-\lam_{L,3} S_2\,l_3\,L'-\lam_{Q,i=1,..3}\,S_1 q_{i}\,Q'\nonumber\\
 & &+M_L L L'+M_E E E'+M_Q Q Q' +\mu_S S_1 S_2\,,\label{superpot}
\eea
where in the first line of \refeq{superpot} one can read the MSSM contributions, with Yukawa couplings $Y_{e,d,u}$ 
that should be read as $3\times 3$ matrices, and an implied anti-symmetric sum over SU(2) indices. We have noted with $\lam_{E,2,3}$, $\lam_{L,2,3}$, $\lam_{Q,i}$, $\widetilde{Y}_{1,2,l_1,e_1}$ the new Yukawa couplings 
allowed by the symmetries, $\mu_S$ is the $\mu$-term of the superfields $S_1$, $S_2$, and $M_{L,E,Q}$ are VL superpotential mass terms. 
We summarize the quantum numbers of the SM/MSSM fields in Appendix~\ref{app:model}.  
Note that with respect to the MSSM particle content, 
the model introduced in \refeq{superpot} is characterized by one extra up-type quark, one extra down-type quark, two extra leptons of 
electric charge $-1$ and one extra TeV-scale Dirac neutrino, and their antiparticles. 

The terms of the soft SUSY-breaking Lagrangian \textit{beyond} the usual MSSM contributions read
\bea\label{softlagr}
\mathcal{L}&\supset& -\left(m_L^2 |\tilde{L}|^2+m_{L'}^2 |\tilde{L}'|^2+m_E^2 |\tilde{E}|^2+m_{E'}^2 |\tilde{E}'|^2+m_Q^2 |\tilde{Q}|^2+m_{Q'}^2 |\tilde{Q}'|^2+\right.\nonumber\\
 & &\left.+m_{S_1}^2 |S_1|^2+m_{S_2}^2 |S_2|^2\right)-\left(m_{\tilde{l}^{\dag}_1}^2 \tilde{l}_1^{\dag}\tilde{L} +m_{\tilde{e}^{\dag}_R}^2 \tilde{e}_R^{\dag}\tilde{E}+m_{\tilde{l}_1}^2 \tilde{l}_1\tilde{L}' +m_{\tilde{e}_R}^2 \tilde{e}_R\tilde{E}'+\textrm{h.c.} \right)\nonumber\\
 & &-\left(B_{M_L}\,\tilde{L}\tilde{L}'+B_{M_E}\,\tilde{E}\tilde{E}'+B_{M_Q}\,\tilde{Q}\tilde{Q}'+B_{\mu_S}\,\tilde{S}_1\tilde{S}_2+\textrm{h.c.}\right)\nonumber\\
 & &+\left(a_{S_2}^{(2)}S_2\tilde{E}'\tilde{\mu}_R+a_{S_1}^{(3)}S_1\tilde{E}'\tilde{\tau}_R+a_{S_1}^{(2)}S_1\tilde{L}'\tilde{\l}_2+a_{S_2}^{(3)}S_2\tilde{L}'\tilde{l}_3+a_{S_1}^{Q}S_1\tilde{Q}'\tilde{q}\right.\nonumber\\
 & &\left.+a_{Y_1} H_d \tilde{L}\tilde{E}+a_{Y_2}H_u\tilde{L}'\tilde{E}'+a_{Y_{l_1}}H_d \tilde{l}_1\tilde{E}+a_{Y_{e_1}}H_d \tilde{L}\tilde{e}_R+\textrm{h.c.}\right)\,,
\eea
where the coefficients of the terms in \refeq{softlagr} represent new soft masses, B-terms, and A-terms, with self-explanatory meaning of the symbols.  
One finds additional states in the sfermion spectrum with respect to the MSSM:
4 extra squarks, 4 extra charged sleptons, and 2 sneutrinos. 

Finally, there are 4 real degrees of freedom in the 
2 complex SM singlets $S_1$ and $S_2$, one of which is transferred to the massive $Z'$ gauge boson, whereas 
the remaining ones give rise to 2 additional neutral ``Higgs" fields, $h'$, $H'$,   
and one pseudoscalar, $A'$. The gaugino content is enriched by 1 Majorana-like $Z'$-\textit{ino}, $\widetilde{Z}'$, and, after 
U(1)$_X$ breaking, 2 additional Majorana \textit{singlinos}, $\widetilde{S}_1$ and  $\widetilde{S}_2$.
The tadpole equations of the scalar fields beyond the MSSM ones, and the scalar mass matrix diagonalization, are shown in Appendix~\ref{app:model}. 
The parameter of greatest impact in the new scalar sector is the ratio of the vevs of the scalar fields $S_1$ and $S_2$, $\tanb_S\equiv v_1/v_2$.

As was mentioned above, in this work we refrain from investigating GUT completions, or other possible UV extensions of the 
model in \refeq{superpot}. We rather focus on the phenomenological and experimental signatures that set it apart from the 
MSSM and non-SUSY versions of $Z'$ models.
For this reason we make throughout the paper the simplifying assumption, often invoked in the literature, that the kinetic mixing of the U(1)$_Y$ and U(1)$_X$ gauge bosons is negligible at the scale of interest for the experimental signatures.
Indeed, radiative corrections arising below the scale at which the VL superfields $Q$ and $Q'$ decouple do in principle generate a 
non-zero kinetic mixing $\epsilon'$ as\cite{Holdom:1985ag} 
\be
\epsilon' \sim\frac{g_1 g_X}{48\pi^2}\ln\left(\frac{M_Q^2}{\mu^2}\right),
\ee
where $g_1$  and $g_X$ are the U(1)$_Y$ and U(1)$_X$ gauge couplings, respectively, and $\mu$ is the renormalization scale. 
This quantity, however, never exceeds typical values of the order of $\sim 10^{-3}$ for $M_Q$ and $g_X$ ranges considered in this study and can be neglected in a first approximation. 
Incidentally, note that $\epsilon'\approx 0$ also implies there is no tree-level mixing between the MSSM and the new sector fields neither in the Higgs sector, nor in the neutralino mass matrices. As we shall see, this feature has important phenomenological consequences when it comes to the LHC constraints on the model.

We also do not attempt to embed the model in a theory of flavor, which might eventually justify  
the structure of the Yukawa couplings emerging from the phenomenological analysis. Effectively, this means we delegate the theoretical 
explanation of whatever texture is favored by the data to the specifics of the unknown UV completion. 
In this spirit we further reduce the number of input parameters by imposing that the following parameters are negligible: 
$\lam_{E,3}\approx\lam_{L,3}\approx\lam_{Q,1}\approx\widetilde{Y}_{l_1}\approx\widetilde{Y}_{e_1}\approx0$.
Note that the latter assumption also prevents our model from generating flavor-changing neutral currents among the MSSM 
fields, which are strongly constrained experimentally. 
 
In conclusion, the only way the MSSM and the extra sectors communicate with each other at tree-level, 
apart from a direct coupling of \zp\ to muons and taus,  is through the mixing of the new quark and lepton fields with the SM quarks 
and lepton via the vevs of $S_1$ and $S_2$. The size of this mixing, together with the value of the gauge coupling $g_X$, are the main parameters of interest while discussing the phenomenological properties of the model.\footnote{At the one-loop level, the VL quarks and leptons further mix the two sectors. These effects are included in our numerical results.}
We present explicitly the tree-level fermion mixing matrices, charged slepton mass matrix, neutralino mass matrix and 
new Higgs states of the model in Appendix~\ref{app:model}. 

\section{Flavor signatures and the muon \textit{g} -- 2}\label{sec:flav}

We give in what follows a brief overview of the flavor structure of the model, which provides a simultaneous solution 
to the $R_{K/K^{\ast}}$ puzzles and the \gmtwo\ anomaly. We further observe these properties are consistent with expectations of a DM particle close to the TeV scale.

\subsection{LFUV puzzles and \textit{B}-meson mixing}\label{sec:rk}

The solution to the \rk /\rks\ puzzle does not deviate significantly from the well-studied non-SUSY $L_{\mu}-L_{\tau}$ counterpart\cite{Altmannshofer:2014cfa}.  The impact of new physics in $b\to s\,l\,l$ transitions is usually described in a model-independent way by the effective Hamiltonian
\be\label{hameff}
\mathcal{H}_{eff}=-\frac{4G_F}{\sqrt{2}}V_{tb}V_{ts}^*\sum_i(C_iO_i+C_i^{'}O_i^{'})+\textrm{h.c.}\,,
\ee
where $G_F$ is the Fermi constant and $V_{tb}$ and $V_{ts}$ are the elements of the CKM matrix.
Equation~(\ref{hameff}) gives a linear combination of dimension-six operators, $O_i$, and the corresponding Wilson coefficients $C_i$.

Dominant contributions to semi-leptonic decays are given by four-fermion contact operators:
\bea
O_9=\frac{e^2}{16\pi^2}(\bar{s}_L\gamma^\mu b_L)(\bar{l}\gamma_\mu l)\,, &\quad& O_9^{'}=\frac{e^2}{16\pi^2}(\bar{s}_R\gamma^\mu b_R)(\bar{l}\gamma_\mu l)\,,\\
O_{10}=\frac{e^2}{16\pi^2}(\bar{s}_L\gamma^\mu b_L)(\bar{l}\gamma_\mu\gamma_5 l)\,, &\quad& O_{10}^{'}=\frac{e^2}{16\pi^2}(\bar{s}_R\gamma^\mu b_R)(\bar{l}\gamma_\mu\gamma_5 l)\,,
\eea
with $l=e,\mu,\tau$. The impact of scalar operators is negligible once the constraints from \bsmumu\ are taken into account\cite{Alonso:2014csa}. On the other hand, chromomagnetic dipole operator and four-quark operators do not contribute to the LFUV processes.

New physics contributions (denoted by the subscript NP) to Wilson coefficients $C_9$ and $C_{10}$
are constrained by a number of experimental measurements, 
several of which reported anomalies in $b\to s\mu^+\mu^-$ transitions. The most up-to-date global fit\cite{Altmannshofer:2017yso} takes into account the measurement of $B^0\to K^{*0}\mu^+\mu^-$ angular distribution, by the LHCb Collaboration\cite{Aaij:2013qta,Aaij:2015oid} 
and Belle\cite{Abdesselam:2016llu,Wehle:2016yoi}, decay rate deficits in the branching ratios 
$B_s^0\to \phi\mu^+\mu^-$\cite{Aaij:2015esa}, $B^0\to K^0\mu^+\mu^-$, $B^+\to K^+\mu^+\mu^-$, $B^+\to K^{*+}\mu^+\mu^-$\cite{Aaij:2014pli}, as well as determination of the LFUV observables $R_K$\cite{Aaij:2014ora} and $R^*_K$\cite{Aaij:2017vbb}.
The allowed ranges of the coefficients $C_{9,\textrm{NP}}^{\mu}$ and $C_{10,\textrm{NP}}^{\mu}$ at $2\,\sigma$ have emerged to be
\bea
C_{9,\textrm{NP}}^{\mu}&=&[-1.7: -0.6]\,,\label{Wilsonfit9}\\
C_{10,\textrm{NP}}^{\mu}&=&[-0.1: 0.7]\,,\label{Wilsonfit10}
\eea
where the superscript $\mu$ indicates a contribution from the operators with two muons in the final state. Note that, in order to explain LFUV, the equivalent
$C_{9,10,\textrm{NP}}^e$ coefficients for the electron must not show a comparable deviation from the SM. Furthermore, the $C_{9,10}^\prime$ coefficients have been found to be consistent with the null SM expectation.

In a generic model with a tree-level contribution from a heavy non-universal $Z'$ gauge boson,
\be
\mathcal{L}\supset \Delta_L^{bs}\, \bar{s}_L\gamma^{\nu}Z'_{\nu}\,b_L+\frac{1}{2}\Delta^{\mu\mu}_{9}\,\bar{\mu}\gamma^{\nu}Z'_{\nu}\,\mu+\frac{1}{2}\Delta^{\mu\mu}_{10}\,\bar{\mu}\gamma^{\nu}\gamma^5 Z'_{\nu}\,\mu\,,
\ee 
where the contributions to the right-handed quark current are negligible (as is the case in our model), the above-mentioned Wilson coefficients can be expressed as\cite{Altmannshofer:2014rta}
\bea\label{C9generic}
C_{9,\textrm{NP}}^{\mu}&=&-\frac{\Delta_{L}^{bs}\Delta^{\mu\mu}_9}{V_{tb}V_{ts}^{\ast}}\left(\frac{\Lambda_v}{m_{Z'}}\right)^2,   \qquad C_{9,\textrm{NP}}^{\prime \  \mu} =0 \, , \nonumber\\
C_{10,\textrm{NP}}^{\mu}&=&-\frac{\Delta_{L}^{bs}\Delta^{\mu\mu}_{10}}{V_{tb}V_{ts}^{\ast}}\left(\frac{\Lambda_v}{m_{Z'}}\right)^2 \, , \qquad C_{10,\textrm{NP}}^{\prime \  \mu} =0 \, ,
\eea
where $m_{Z'}$ is the mass of the $Z'$ boson, 
\be
\Lambda_v=\left(\frac{\pi}{\sqrt{2}G_F\alpha_{\textrm{em}}}\right)^{1/2}\approx 4.94\tev,
\ee
is the typical effective scale of the new physics, and $\Delta_L^{bs}$, $\Delta^{\mu\mu}_{9,10}\equiv \Delta_R^{\mu\mu}\pm\Delta_L^{\mu\mu}$ are the effective couplings of the 
$Z'$ boson to the SM-like particles. 

The hadronic coupling, $\Delta_L^{bs}$, is subject to strong constraints from the measurement of $B_s$ meson mixing.   
The quantity of interest is\cite{Altmannshofer:2014rta}
\be\label{bsmix}
R_{BB}\equiv\frac{M_{12}}{M_{12,\textrm{SM}}}-1=\frac{\left(246\gev\times\Delta_L^{bs}\right)^2}{m_{Z'}^2}\left[\frac{g_2^2}{16\pi^2}\left(V_{tb}V_{ts}^{\ast}\right)^2 S_0\right]^{-1},
\ee
where $S_0 \approx 2.3$ is a loop function. The bounds from $B_s$ meson mixing have been recently updated to include the latest  Flavour Lattice Averaging Group (FLAG) results\footnote{Note that there is some ongoing disagreement on the precise value of the SM prediction, whose error drives the one of $R_{BB}$. In particular, the UTfit collaboration Summer 2016 results\cite{Bona:2007vi} lead to $R_{BB} = 0.070 \pm 0.088$ while other results (see\cite{DiLuzio:2017fdq}) point instead to a negative $R_{BB}$ with a $1.8 \,\sigma$ tension with the experimental value. We will use the strongest bound from\cite{DiLuzio:2017fdq} in the following.} yielding the limits at $2\sigma$\cite{DiLuzio:2017fdq} 
\be
-0.21 < R_{BB} < 0.014 \ .
\ee

In our model, $Z'$ directly couples with flavor-diagonal strength $g_X$ 
to the muon and tau gauge eigenstates 
and to VL quark doublets $Q$, $Q'$ (we assume all U(1)$_X$ charges $Q_X=\pm 1$, cf.~\refsec{sec:model}).
In order to impose the above constraints, 
the couplings $\Delta_{L}^{bs}$ and $\Delta^{\mu\mu}_{L,R}$ of the $Z'$ boson to the physical $b$ and $s$ quarks and muons 
have to be eventually expressed
in terms of the model's input parameters. To do this, we diagonalize the mass matrices given in Appendix~\ref{app:model} 
by means of two pairs of unitary matrices, $D_L,D_R$ and $D_L',D_R'$, such that 
\be
\mathcal{D}_d=D_L^{\dag} \mathcal{M}_d D_R\,, \quad \mathcal{D}_e=D_L^{\prime \, \dag} \mathcal{M}_e D_R',
\ee
where $\mathcal{M}_d$ and $\mathcal{M}_e$ are, respectively, the down-type quark and charged-lepton mass matrices,  
presented explicitly in Appendix~\ref{app:model}, and $\mathcal{D}_d$ and $\mathcal{D}_e$ are the corresponding diagonal matrices.

By adopting a standard notation for mass eigenstates $d_i, e_i$ in terms of gauge eigenstates $d'_i, e'_i$ -- ordered like in Appendix~\ref{app:model} -- 
so that 
\be
d_{L(R)\,j}=\sum_r D^{\dag}_{L(R)\,jr}d_{L(R)\,r}' \ ,  \quad e_{L(R)\,j}=\sum_r D^{\prime \, \dag}_{L(R)\,jr}e_{L(R)\,r}'\,,\label{diagproc}
\ee
one gets
\bea
\Delta_L^{bs}&=&g_X Q_X\,D^{\dag}_{L\,34}D_{L\,42}\,,\label{bsMI}\\
\Delta^{\mu\mu}_L&=&g_X Q_X\,|D_{L\,22}'|^2 , \\
\Delta^{\mu\mu}_R&=&g_X Q_X\,|D_{R\,22}'|^2\label{muMI}.
\eea
One can easily infer from the form of the down-type quark mass matrix, \refeq{bmatrix}, that the 
right-handed effective coupling of the $b,s$ quarks to the $Z'$ boson, $D_{R\,43}D^{\dag}_{R\,24}$, is zero at tree level.
Also, since we assume $\lam_{E,3}\approx \lam_{L,3}\approx 0$, at the tree level 
the muons and taus do not mix and one can 
safely neglect $D'_{L\,32}\approx D'_{R\,32}\approx 0$.

Equations~(\ref{bsMI})-(\ref{muMI}) are to a very good approximation expressed in terms of the model's input 
parameters as
\bea
\Delta_L^{bs}&\approx& g_X Q_X\frac{\lam_{Q,2}\lam_{Q,3}\,v_1^2}{2 M_Q^2+\left(\lam_{Q,2}^2+\lam_{Q,3}^2\right) v_1^2}, \label{deltaLbs}\\
\Delta_{L(R)}^{\mu\mu}&\approx& g_X Q_X \frac{2 M_{L(E)}^2}{2 M_{L(E)}^2+\lam_{L(E),2}^2 v_{1(2)}^2}\,. 
\eea

\begin{figure}[t]
\centering
\subfloat[]{%
\includegraphics[width=0.47\textwidth]{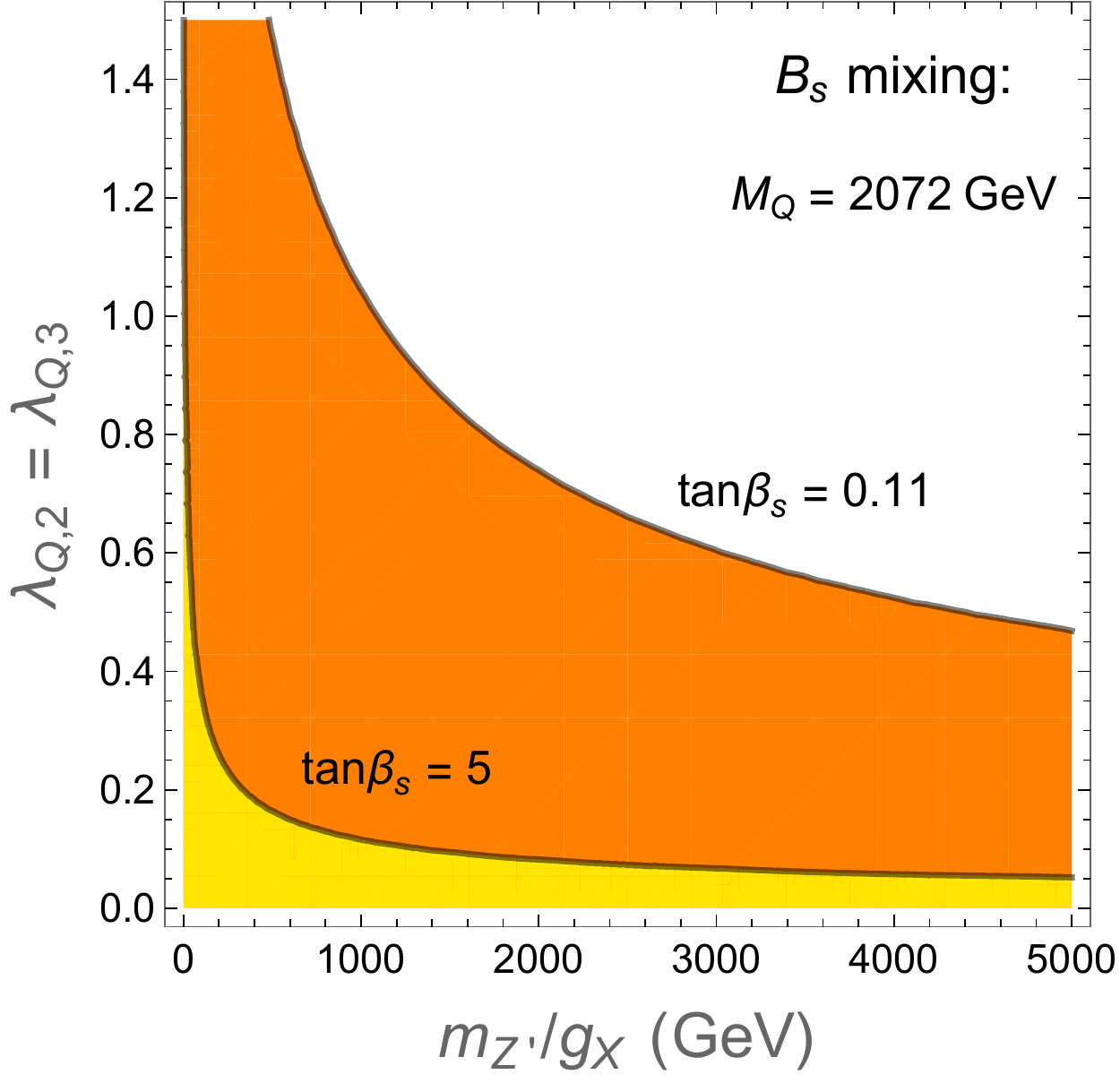}
}%
\hspace{0.02\textwidth}
\subfloat[]{%
\includegraphics[width=0.47\textwidth]{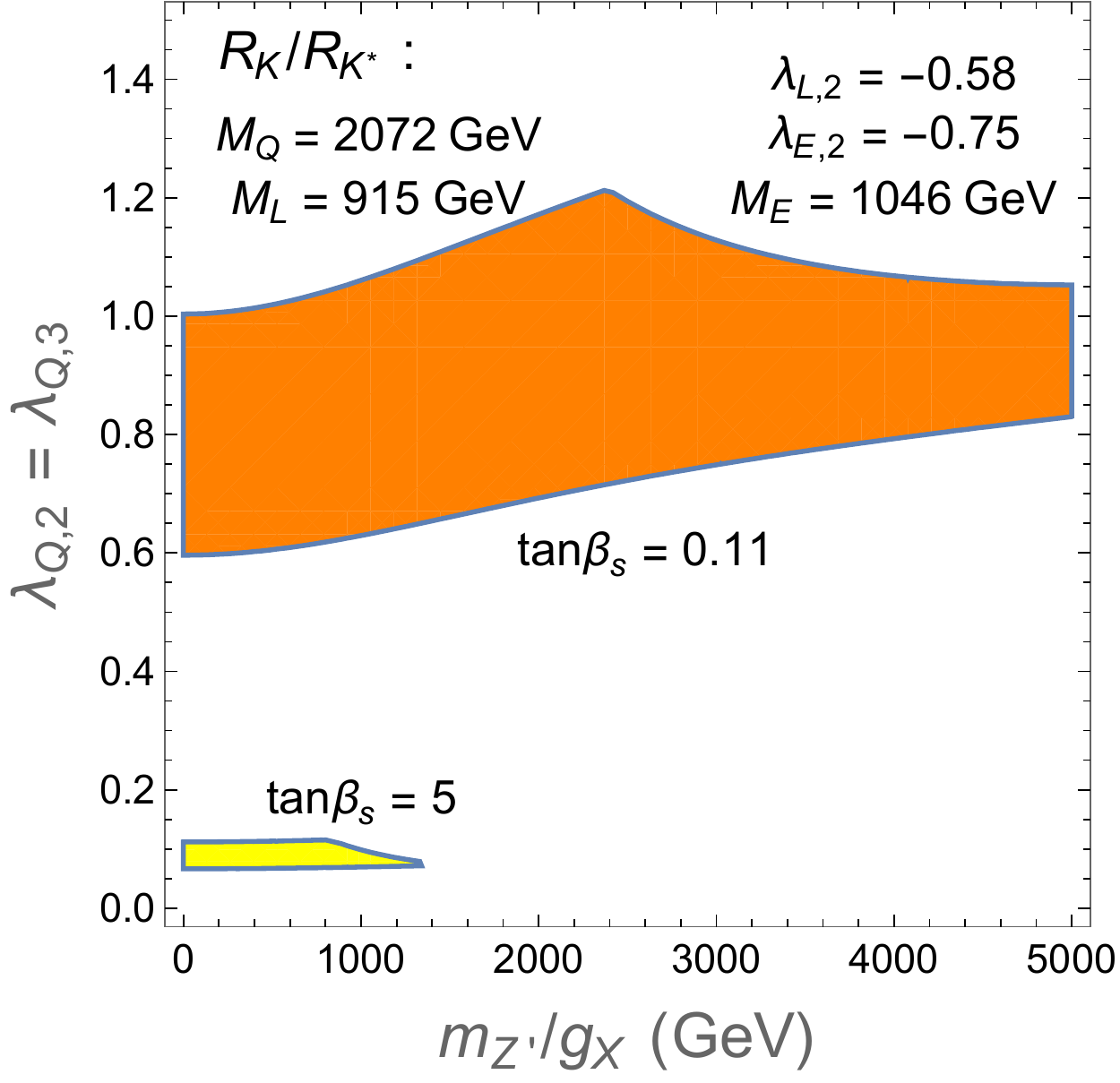}
}%
\caption{(a) The parameter space consistent with constraints from $B_s$ mixing in the ($m_{Z'}/g_Z$, $\lam_{Q,2}=\lam_{Q,3}$) plane 
for two choices of $\tanb_S$:
$\tanb_S=0.11$ in orange and $\tanb_S=5$ in yellow. 
(b) The impact of the $R_{K}/R_{K^{\ast}}$ constraints on the $C_{9,\textrm{NP}}^{\mu}$ and $C_{10,\textrm{NP}}^{\mu}$ Wilson coefficients (Eqs.~(\ref{Wilsonfit9}) and (\ref{Wilsonfit10}))
in the same plane, for the same two choices of $\tanb_S$.}
\label{fig:flavor}
\end{figure}

In \reffig{fig:flavor}(a) we show in the ($m_{Z'}/g_X$, $\lam_{Q,2}=\lam_{Q,3}$) plane 
the parameter space consistent at the $2\,\sigma$ level 
with the $B_s$ mixing measurement\cite{DiLuzio:2017fdq}.
We choose two representative values of $\tanb_S$:
$\tanb_S=0.11$ (orange parameter space) and $\tanb_S=5$ (yellow). 
Note that the parameter space is much more constrained for $\tanb_S>1$, 
due to the fact that the mixing between the U(1)$_X$-charged 
quark $D_L$ and the SM down-type quarks is directly proportional to the $v_1$ vev only. 
This is a feature that will affect the bounds on the quark mixing sector throughout this work.

In \reffig{fig:flavor}(b) we show the impact of the $R_{K}/R_{K^{\ast}}$ constraint
on the $C_{9,\textrm{NP}}^{\mu}$ and $C_{10,\textrm{NP}}^{\mu}$ Wilson coefficients -- i.e. Eqs.~(\ref{Wilsonfit9}) and (\ref{Wilsonfit10}) are satisfied within the colored regions --
in the ($m_{Z'}/g_X$, $\lam_{Q,2}=\lam_{Q,3}$) plane, for the same two choices of $\tanb_S$. 
Again, one can see that much larger parameter space is available when $\tanb_S\ll 1$. 

\subsection{Muon \textit{g} -- 2 and connections to dark matter\label{sec:gmtwo}}

We recall that, once recent estimates of the hadronic and light-by-light scattering uncertainties are taken into account, 
the measured value of the anomalous magnetic moment of the muon, \gmtwo, shows a $\sim3.5\,\sigma$ discrepancy with the SM:
$\deltagmtwomu=(27.4\pm 7.6)\times 10^{-10}$\cite{Davier:2016iru}. (Another recent review of the hadronic vacuum polarization 
contributions and uncertainties\cite{Jegerlehner:2017lbd} yields an even more convincing $\deltagmtwomu = (31.3 \pm 7.7 ) \times 10^{-10}$.)
In our model one observes several contributions to the calculation of \gmtwo\ in addition to those of the MSSM.
These extra terms can allow a good fit to the \gmtwo\ anomaly within a SUSY spectrum of broader mass range than in the MSSM.

At the one-loop level, the dominant new-physics contribution to \gmtwo\ 
can be roughly subdivided into two categories\cite{Leveille:1977rc,Moore:1984eg}. 
First, there are those due to  
neutral scalar fields (collectively indicated by $S^0$) and electrically charged fermions 
(collectively indicated by $E^{\pm}$), which read  
\be\label{chiralint}
\deltagmtwomu=\frac{1}{16\pi^2}\sum_{S^0,E^{\pm}}
\left[\frac{m_{\mu}^2}{m_{S^0}^2}\left(|c_L|^2+|c_R|^2\right)\mathcal{F}_1(r)
+2\,\frac{m_{\mu} m_{E^{\pm}}}{m_{S^0}^2}\,\Re(c_L c_R^{\ast})\,\mathcal{F}_2(r)\right]\,,
\ee
where $m_{\mu}$ is the muon mass, $c_L$ and $c_R$ are the direct couplings of the BSM fields to the left- and right-chiral states of the muon, respectively, 
$r\equiv (m_{E^{\pm}}/m_{S^0})^2$, and the loop functions are given by 
\bea
\mathcal{F}_1(r)&\equiv&\frac{2+3r-6r^2+r^3+6r\ln r}{6\,(r-1)^4} ,   \\
\mathcal{F}_2(r)&\equiv&\frac{3-4r+r^2+2\ln r}{2\,(r-1)^3} \,. 
\eea
Secondly, there are those from charged scalars $S^\pm$ and neutral fermions $E^0$ in the loop: 
\be\label{chargescal}
\deltagmtwomu=\frac{1}{16\pi^2}\sum_{S^{\pm},E^0}\left[-\frac{m_{\mu}^2}{m_{S^{\pm}}^2}\left(|c_L|^2+|c_R|^2\right)\mathcal{G}_1(r)
+\frac{m_{\mu} m_{E^0}}{m_{S^{\pm}}^2}\,\Re(c_L c_R^{\ast})\,\mathcal{G}_2(r)\right]\,, 
\ee
where 
\bea
\mathcal{G}_1(r)&\equiv&\frac{1-6 r+3 r^2+2 r^3-6 r^2\ln r}{6(r-1)^4} ,   \\
\mathcal{G}_2(r)&\equiv&\frac{-1+r^2-2 r\ln r}{(r-1)^3}\,.  
\eea
Note that \refeq{chargescal} is negative when the fields of the new physics couple to just one of the chiral states of the muon, so that either  $c_L$ or $c_R$ is equal zero, 
but it can become positive for 
$\Re(c_L c_R^{\ast})\neq 0$. 

In addition to these two groups, one finds in our model contributions from the exchange of neutral vector fields (like the $Z'$) 
and charged fermions in the loop, but they are subdominant in our model and we do not discuss them in what follows.  

As is well known, in the presence of chirality-flip interactions of the new physics states with the muon 
(the term in $c_L c_R^{\ast}$ in Eqs.~(\ref{chiralint}) and (\ref{chargescal})) the calculation of \gmtwo\ receives a welcome boost from  
BSM contributions proportional to the bare mass of the new VL fermions, $m_E$\cite{Grifols:1982vx,Czarnecki:2001pv,Kowalska:2017iqv}.
However, it is also well known (see, e.g.,\cite{Fukushima:2014yia}) that these contributions introduce sizable, albeit cut-off independent, loop contributions in the calculation of the muon mass, 
as the latter is not in this case protected 
by chiral symmetry. One finds
\be
\delta m_{\mu}\sim \frac{m_E}{16\pi^2}\Re\left(c_L c_R^{\ast}\right)\log\left(\frac{m_{E_2}}{m_{E_1}}\right)\sim \frac{m_E}{16\pi^2}\Re\left(c_L c_R^{\ast}\right)\log\left(\frac{m_{S_2}}{m_{S_1}}\right)\,
\ee
for any pair of mixing fermions $m_{E_{1,2}}$ (or scalars $m_{S_{1,2}}$) in the loop.
This problem is generally dealt with by adjusting the value of the muon Yukawa coupling (which is not strongly constrained experimentally) 
until the large loop corrections are canceled point by point in the parameter space. 
We have designed our numerical scans so to have this procedure automatically carried out. 
The benchmark points we present in \refsec{sec:bench} are
all characterized by the physical muon mass within $\sim30\%$ of the experimental value, where we have left some room for the unknown uncertainties of higher-order loop contributions.  

\begin{figure}[t]
\centering
\subfloat[]{
\label{fig:a}
\includegraphics[width=0.4\textwidth]{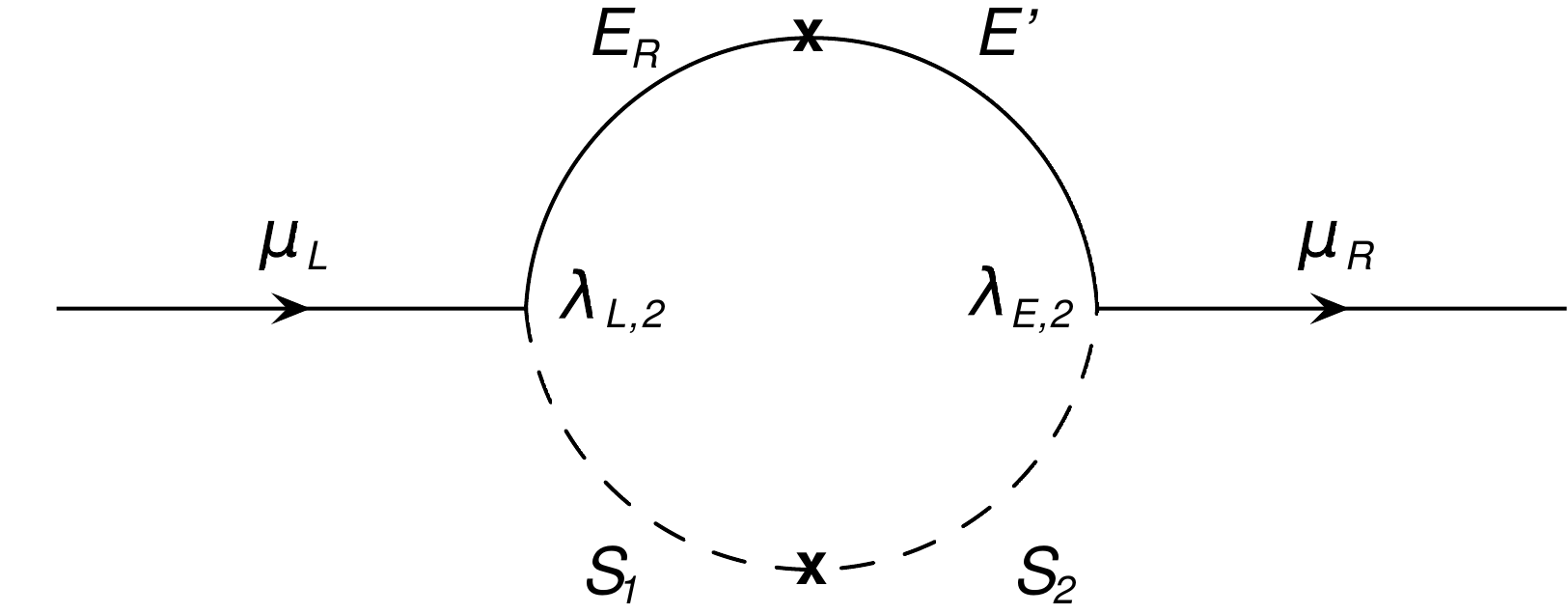}
}
\\
\hspace{1cm}
\subfloat[]{
\label{fig:b}
\includegraphics[width=0.4\textwidth]{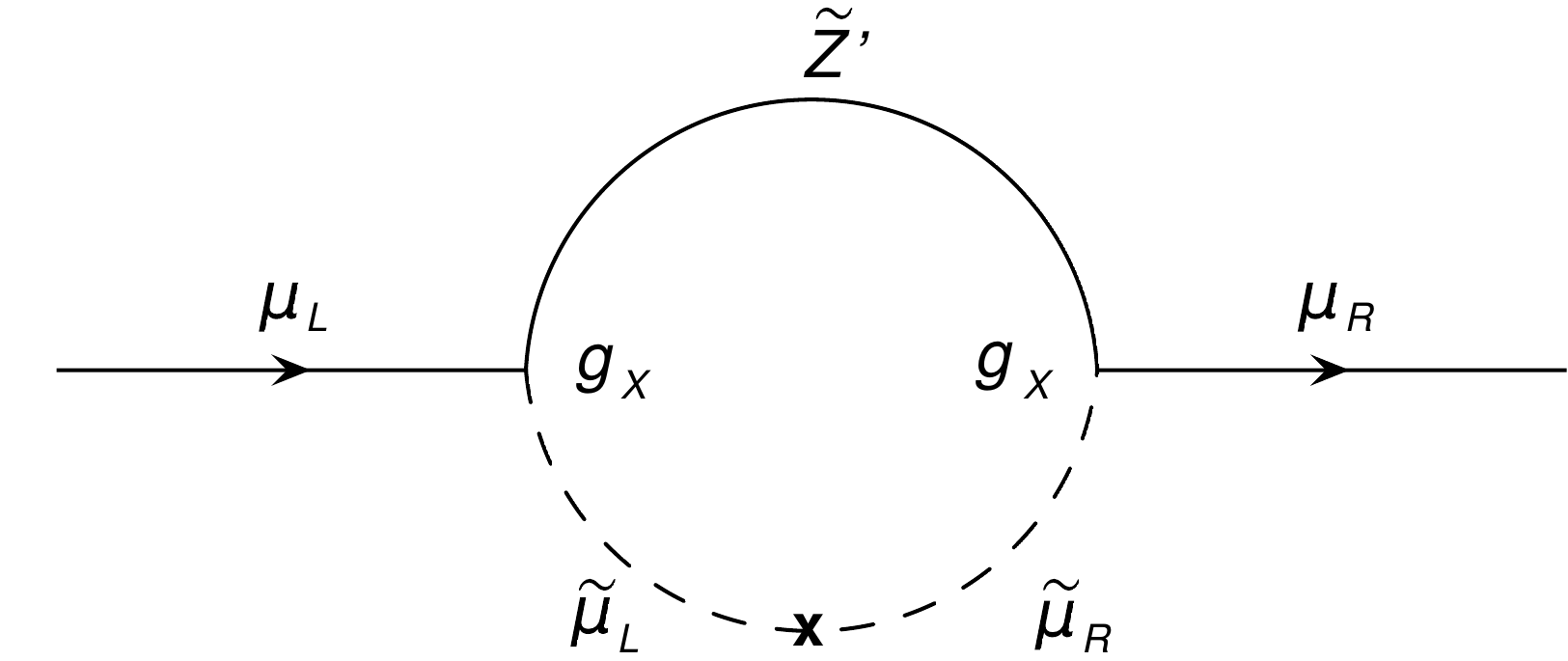}
}
\hspace{1cm}
\subfloat[]{
\label{fig:c}
\includegraphics[width=0.4\textwidth]{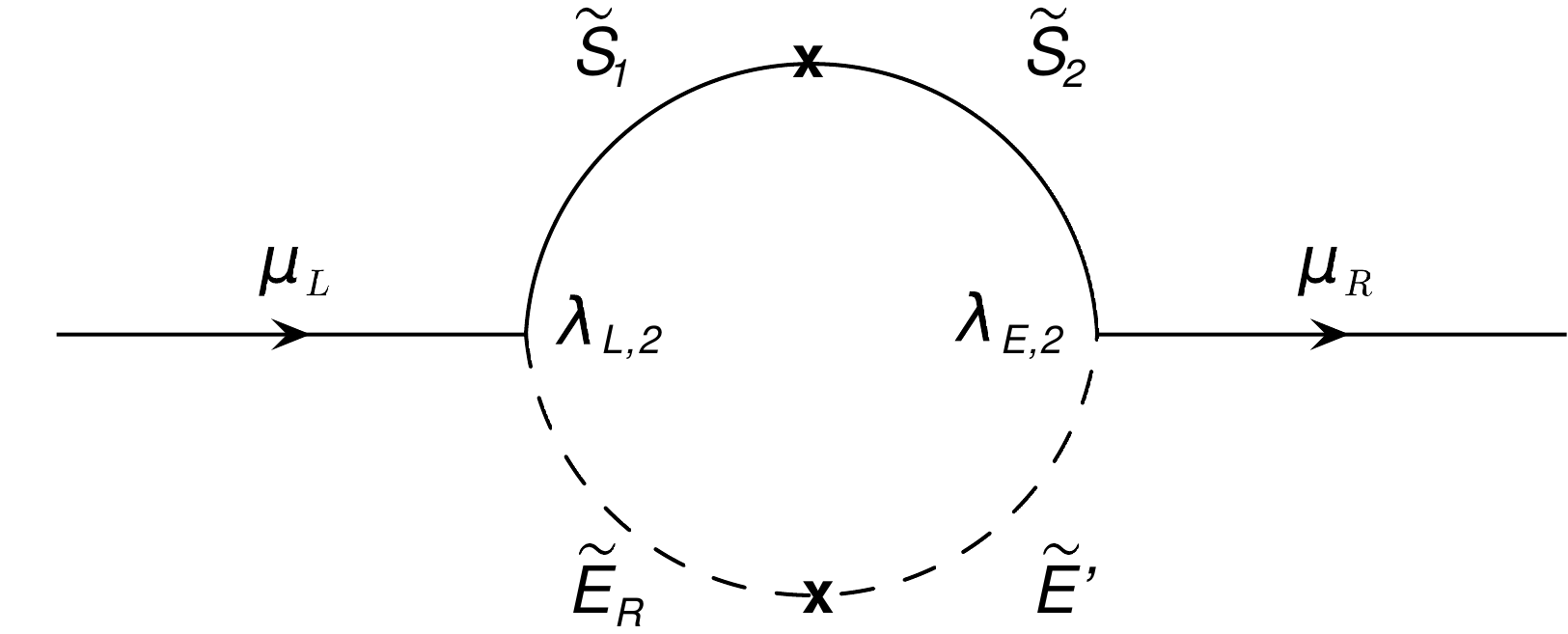}
}
\caption{(a) The contribution to \gmtwo\ from the singlet scalars and VL lepton fields. 
(b), (c) Examples of contributions from neutralino/slepton loops that \textit{cannot} be found in the MSSM. 
In all cases it is implied the photon line is attached to the electrically charged particle in the loop.}
\label{fig:g2diagrams}
\end{figure}

As pertains to the contributions to \gmtwo, in the regular MSSM\cite{Moroi:1995yh,Martin:2001st} these are generated by chargino/sneutrino loops, corresponding to \refeq{chiralint}, and 
neutralino/smuon loops, corresponding to \refeq{chargescal}. 
In the model described here, though, there exist important extra terms.

A first class of these diagrams are 
induced by the neutral (pseudo)scalars $h'$, $H'$, $A'$, and the charged heavy leptons corresponding to 
$L'$ and $E'$, as shown in \reffig{fig:g2diagrams}(a).
By parameterizing the mixing of the new scalar and pseudoscalar fields via the angles $\alpha'$ and $\alpha''$, respectively, as is done in Appendix~\ref{app:model}, 
one can express the couplings of the new scalars to the left- and right-chiral components of the muon and the 5 
physical charged leptons, labeled here in order of increasing mass, $I=1,..,5$: 
\bea\label{clscal}
c_L^{I\,h'}&=&\lam_{E,2} \cos\alpha' D'^{\dag}_{L\,25}D'^{\dag}_{R\,I2}+\lam_{L,2} \left(-\sin\alpha'\right) D'^{\dag}_{L\,22}D'^{\dag}_{R\,I4}\\
c_L^{I\,H'}&=&\lam_{E,2} \sin\alpha' D'^{\dag}_{L\,25}D'^{\dag}_{R\,I2}+\lam_{L,2} \cos\alpha' D'^{\dag}_{L\,22}D'^{\dag}_{R\,I4}\\
c_L^{I\,A'}&=&i\left(\lam_{E,2} \sin\alpha'' D'^{\dag}_{L\,25}D'^{\dag}_{R\,I2}+\lam_{L,2} \cos\alpha'' D'^{\dag}_{L\,22}D'^{\dag}_{R\,I4}\right),
\eea
\bea
c_R^{I\,h'}&=&\lam_{E,2} \cos\alpha' D'^{\dag}_{R\,22}D'^{\dag}_{L\,I5}+\lam_{L,2} \left(-\sin\alpha'\right) D'^{\dag}_{R\,24}D'^{\dag}_{L\,I2}\\
c_R^{I\,H'}&=&\lam_{E,2} \sin\alpha' D'^{\dag}_{R\,22}D'^{\dag}_{L\,I5}+\lam_{L,2} \cos\alpha' D'^{\dag}_{R\,24}D'^{\dag}_{L\,I2}\\
c_R^{I\,A'}&=&i\left(\lam_{E,2} \sin\alpha'' D'^{\dag}_{R\,22}D'^{\dag}_{L\,I5}+\lam_{L,2} \cos\alpha'' D'^{\dag}_{R\,24}D'^{\dag}_{L\,I2}\right),\label{crscal}
\eea  
where the fermion mixing matrices $D^{\prime \, \dag}_{L,R}$ are defined as in \refeq{diagproc}.
Our numerical analysis is based on a full one-loop numerical calculation, 
which takes into account all of the diagrams and couplings contributing
to the measured observable. It is however worth briefly taking a look at a parametric approximation for the dominant 
contributions to \gmtwo. 

In the limit $\tanb\gg 1$, 
the contribution from one of the new scalar fields, say $H'$, and VL leptons reads 
\begin{multline}
\deltagmtwomu^{(H')} \sim \frac{m_{\mu}\widetilde{Y}_2 v_u}{16\pi^2 m_{H'}^2}\,\lam_{L,2}\lam_{E,2}\,\sin 2\alpha'\\
\times\left[\frac{m_{e_5}\max\left(M_L,M_E\right)}{m_{e_5}^2-m_{e_4}^2}\mathcal{F}_2(r_5)-\frac{m_{e_4}\min\left(M_L,M_E\right)}{m_{e_5}^2-m_{e_4}^2}\mathcal{F}_2(r_4)\right],\label{g2appr}
\end{multline}
where $m_{e_4}, m_{e_5}$ are the masses of the heaviest charged leptons in the spectra and $r_i=m_{e_i}^2/m_{H'}^2$.

\begin{figure}[t]
\centering
\subfloat[]{%
\includegraphics[width=0.47\textwidth]{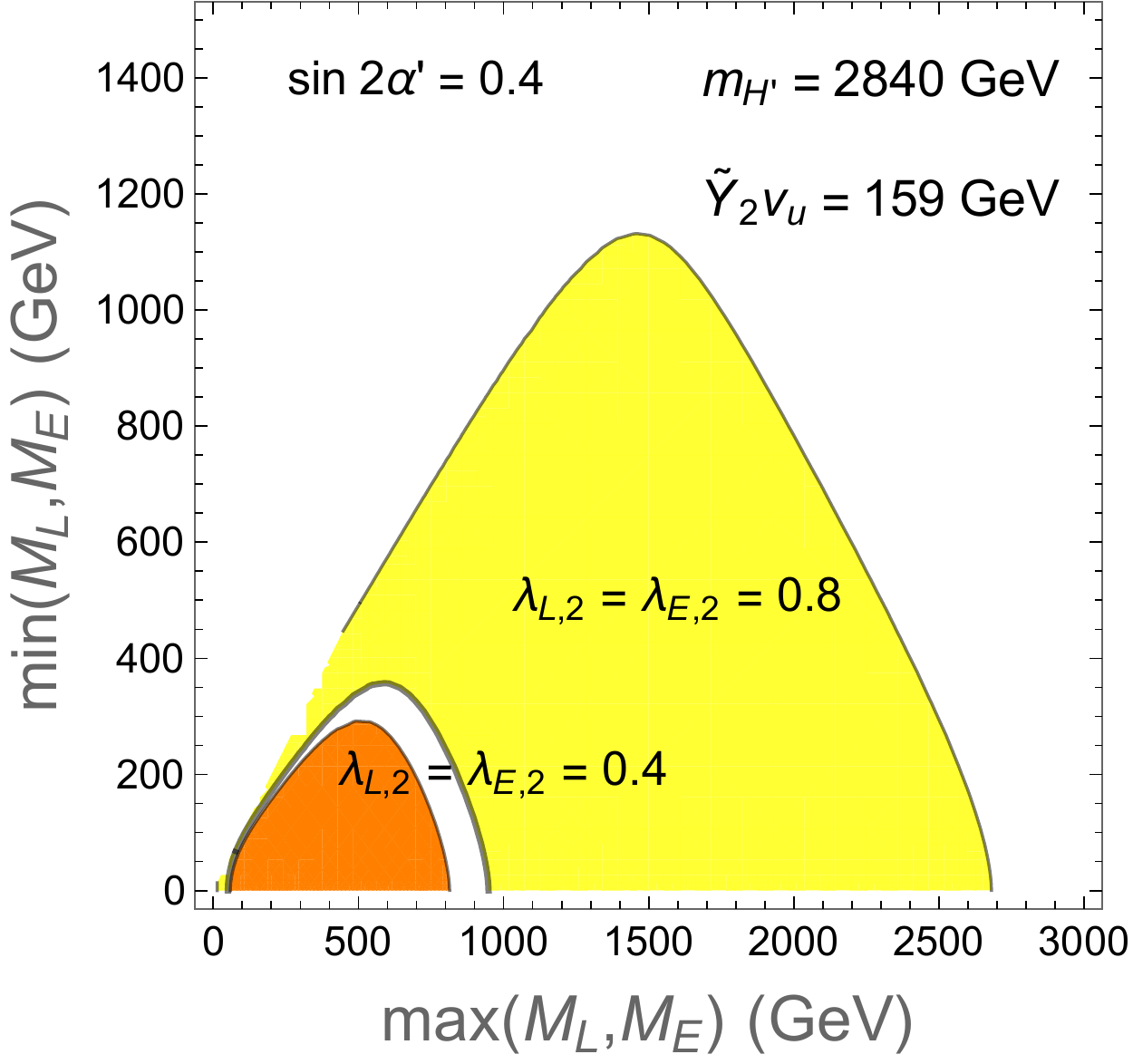}
}%
\hspace{0.02\textwidth}
\subfloat[]{%
\includegraphics[width=0.455\textwidth]{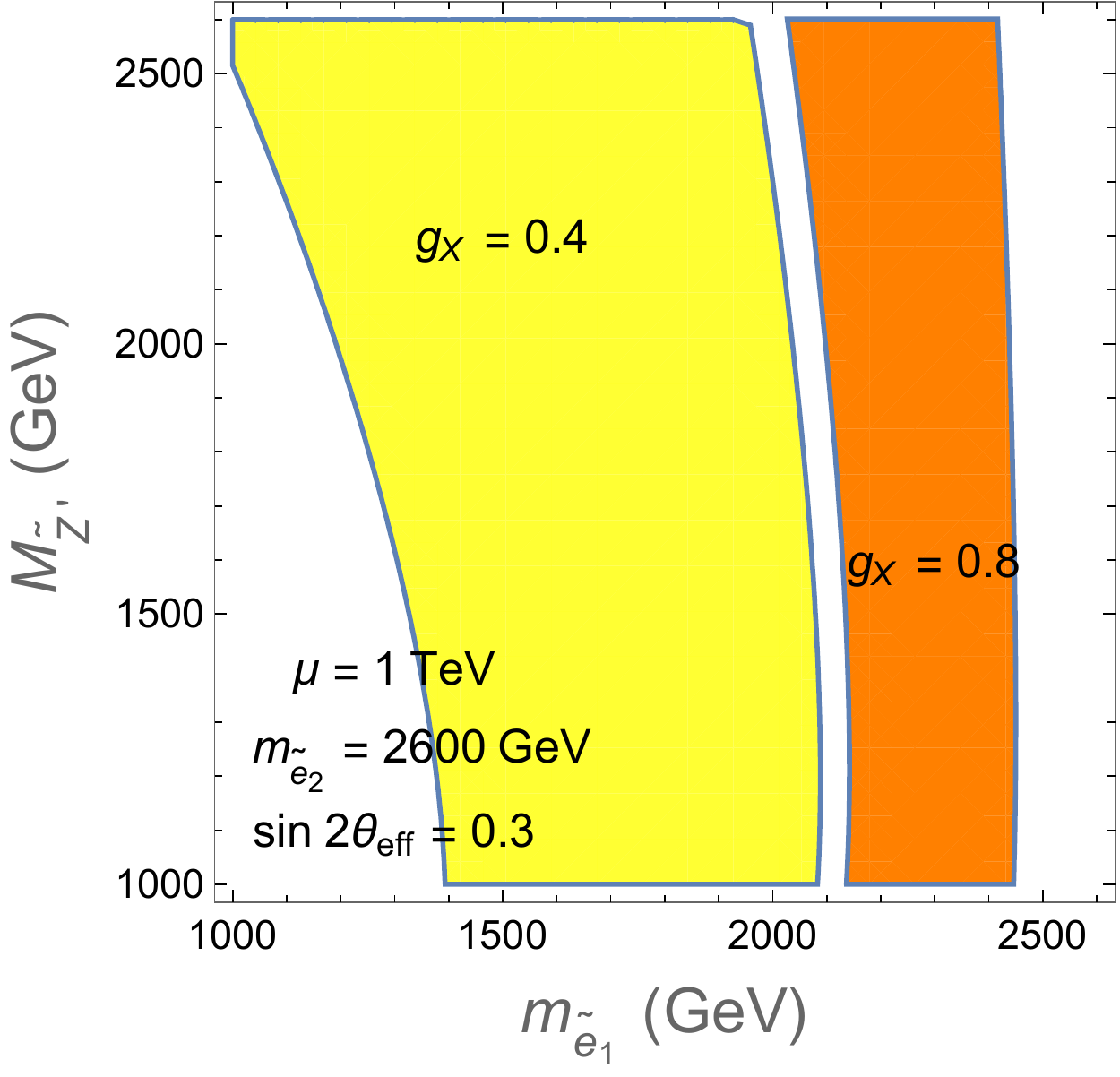}
}%
\caption{(a) The region of the ($M_L$, $M_E$) plane yielding \deltagmtwomu\ in agreement with the experimental measurement at $2\,\sigma$, if 
only the contribution of \refeq{g2appr} is taken into account. In orange, $\lam_{L,2}=\lam_{E,2}=0.4$\,; in yellow, $\lam_{L,2}=\lam_{E,2}=0.8$\,.
The scalar mass and mixing parameters are fixed to typical values featured in the legend. (b) In the ($m_{\tilde{e}_1}$, $M_{\tilde{Z}'}$) plane 
the parameter space in agreement with the measurement of \deltagmtwomu\ at the $2\,\sigma$ level, if only the contribution of \refeq{g2neutapp} is considered. 
In yellow $g_X=0.4$ and in orange $g_X=0.8$. Other relevant parameters are fixed to the values given in the legend. }
\label{fig:g2scalar}
\end{figure}

We plot in \reffig{fig:g2scalar}(a) the parameter space regions yielding \deltagmtwomu\ within $2\,\sigma$ of the measured value
using only the contribution given by \refeq{g2appr}.
The figure is presented in the plane of the bare superpotential VL lepton mass terms, ($M_L$, $M_E$), 
for two choices of new Yukawa couplings, $\lam_{L,2}=\lam_{E,2}=0.4$ (orange region) 
and $\lam_{L,2}=\lam_{E,2}=0.8$ (yellow region). A few other parameters of relevance are fixed as in the legend.   
One can see that the contribution from \refeq{g2appr} becomes larger 
when the separation between the bare mass parameters is made large ($M_L\gg M_E$ or vice versa), 
and/or when lepton mixing due to the MSSM Higgs vev is sizable.
The reader should be aware, on the other hand, that contributions to \deltagmtwomu\ 
from other (pseudo)scalars ($h'$, $ A'$) tend to cancel the contributions of 
\refeq{g2appr}, as at least one of them always enters with opposite sign. 
In order to reduce the cancellation one is thus bound to consider
parameter space characterized by $m_{H'}\gg m_{h'}$, and the mixing angles relative to $H'$, $A'$ should be preferably such that
$\sin\alpha''\approx \sin\alpha'$, $\cos\alpha''\approx \cos\alpha'$.

It is important to note that contributions to \deltagmtwomu\ 
due to new scalar/VL lepton loops, like \refeq{g2appr}, do not depend on the parameters of the neutralino sector 
and can in principle be present independently of the specific features of the SUSY spectrum. 
A consequence of this is that in principle, by means of these contributions alone, the model is able to
resolve the \gmtwo\ anomaly for much broader range of DM mass values and properties than in the MSSM. 

There is, however, a second class of important diagrams that contribute to \gmtwo. They involve combinations of the 7 neutralinos 
and 10 sleptons of the model, see for example \reffig{fig:g2diagrams}(b) and \reffig{fig:g2diagrams}(c).
After ordering the neutralinos and sleptons by their increasing mass, $I=1,..,7$,  $K=1,..,10$, one can write the full expression for the couplings to the muon chiral states, 
\bea
c_L^{IK}&=&-\sqrt{2}g_1\left(y_{e_L}D'^{\dag}_{L\,22}R^n_{I1}R^s_{K3}+y_{E_L}D'^{\dag}_{L\,24}R^n_{I1}R^s_{K7}+y_{E'}D'^{\dag}_{L\,25}R^n_{I1}R^s_{K9}\right)\nonumber\\
 &-&\sqrt{2}g_2\left(T^3_{e_L} D'^{\dag}_{L\,22}R^n_{I2}R^s_{K3}+T^3_{E_L} D'^{\dag}_{L\,24}R^n_{I2}R^s_{K7}\right)-Y_{e,22}D'^{\dag}_{L\,22}R^n_{I3}R^s_{K4}\nonumber\\
 &-&\widetilde{Y}_1 D'^{\dag}_{L\,24}R^n_{I3}R^s_{K10}-\widetilde{Y}_2 D'^{\dag}_{L\,25}R^n_{I4}R^s_{K8}\nonumber\\
 &-&\sqrt{2}g_X Q_X D'^{\dag}_{L\,22}R^n_{I5}R^s_{K3}-\lam_{L,2} D'^{\dag}_{L\,22}R^n_{I6}R^s_{K8}+\lam_{E,2}D'^{\dag}_{L\,25}R^n_{I7}R^s_{K4}\,;\label{g2neut0}\\
c_R^{IK}&=&-\sqrt{2}g_1\left(y_{e_R}D'^{\dag}_{R\,22}R^n_{I1}R^s_{K4}+y_{E_R}D'^{\dag}_{R\,24}R^n_{I1}R^s_{K8}+y_{E}D'^{\dag}_{R\,25}R^n_{I1}R^s_{K\,10}\right)\nonumber\\
 &-&\sqrt{2}g_2 T^3_{E_R} D'^{\dag}_{R\,24}R^n_{I2}R^s_{K8}-Y_{e,22}D'^{\dag}_{R\,22}R^n_{I3}R^s_{K3}\nonumber\\
 &-&\widetilde{Y}_1 D'^{\dag}_{R\,25}R^n_{I3}R^s_{K7}-\widetilde{Y}_2 D'^{\dag}_{R\,24}R^n_{I4}R^s_{K9}\nonumber\\
 &-&\sqrt{2}g_X Q_X D'^{\dag}_{R\,22}R^n_{I5}R^s_{K4}-\lam_{L,2} D'^{\dag}_{R\,24}R^n_{I6}R^s_{K3}+\lam_{E,2}D'^{\dag}_{R\,22}R^n_{I7}R^s_{K9}\,,\label{g2neut}
\eea
in terms of the lepton, slepton, and neutralino mixing matrices defined as in Appendix~\ref{app:model}, gauge couplings $g_1,g_2,g_X$, isospin $T^3$, and hypercharge assignment $y$.

We point out that for this second class of important contributions we expect effects that largely extend the parameter space consistent with the measurement of \gmtwo\ with 
respect to the MSSM. Let us consider, for example, the well-known MSSM case of a nearly pure higgsino DM particle with a mass of $\sim 1\tev$. 
In the MSSM neutralino/slepton loop contributions to \gmtwo\ for a $\sim 1\tev$ higgsino typically involve the product of one of the gauge couplings $g_1,g_2$ and the small 
muon Yukawa coupling, and are suppressed by the negligible $R_{12}^n (R_{13}^n)^{\dag}$ mixing. Conversely, Eqs.~(\ref{g2neut0}) and (\ref{g2neut}) show that in the U(1)-extended model 
one can rely on the product of sizable gauge and/or Yukawa couplings ($2 g_X^2$, $\lam_{L,2}\lam_{E,2}$, etc.) and large mixing angles to boost the calculated value of \gmtwo.

To guide the eye, 
we explicitly write down a parametric approximation of the contribution to \deltagmtwomu\ due to the $Z'$-ino and the two lightest sleptons in the spectrum, cf. \reffig{fig:g2diagrams}(b). 
One gets 
\be\label{g2neutapp}
\deltagmtwomu^{(\tilde{Z'})}\simeq \frac{g_X^2}{16\pi^2}m_{\mu}M_{\tilde{Z}'}\sin 2\theta_{\textrm{eff}}\left[\frac{\mathcal{G}_2(t_1)}{m_{\tilde{e}_1}^2}-\frac{\mathcal{G}_2(t_2)}{m_{\tilde{e}_2}^2}\right],
\ee
in terms of an effective slepton mixing angle, $\theta_{\textrm{eff}}$, and
where we define $t_i=M_{\tilde{Z}'}^2/m_{\tilde{e}_i}^2$ as the squared ratio of the $Z'$-ino mass to the $i$-th slepton mass. Typical values for 
$\sin 2\theta_{\textrm{eff}}$ range in the ballpark of $0.1-0.5$, and its parametric expansion strongly depends on the particular chosen point in the parameter space. When the 
lightest sleptons are nearly pure smuon gauge eigenstates one gets, for example, 
$\sin 2\theta_{\textrm{eff}}\approx 2\,Y_{e,22}\,v_u (\mu\tanb-A_{\mu})/(m_{\tilde{e}_2}^2-m_{\tilde{e}_1}^2)$ .

We show in \reffig{fig:g2scalar}(b) the parameter space in the ($m_{\tilde{e}_1}$, $M_{\tilde{Z}'}$) plane 
in agreement with the measured \deltagmtwomu\ at the $2\sigma$ level,
and with a $\sim 1\tev$ higgsino DM candidate. We plot exclusively the contribution due to \refeq{g2neutapp}, with $\sin 2\theta_{\textrm{eff}}=0.3$. 
The region in yellow is obtained when $g_X=0.4$, the region in orange when $g_X=0.8$, and the other relevant parameters are fixed to the values given in the legend.
One can then see that the parameter space is much extended with respect to the MSSM\cite{Kowalska:2015zja}.

To conclude this section, we point out that in the most realistic cases the contribution to \gmtwo\ is given by a combination of all of these effects, 
which can sum and subtract from each other in a complicated way, depending on the parameter space region at hand. 
In particular, in the benchmark points we present is \refsec{sec:bench}, \deltagmtwomu\ is obtained in agreement with the measured value always as a result of the 
combination of many different diagrams, without one single mechanism playing a dominant role. 

\section{Numerical results and benchmark scenarios\label{sec:bench}}

In this section we present benchmark scenarios that fit the flavor constraints discussed in \refsec{sec:flav} and 
are representative of different interesting parameter space regions of the model. We also analyze their DM properties 
and give an overview of the typical LHC bounds.

\subsection{Benchmark points}

Each benchmark point was generated with the numerical package \texttt{SARAH~v.4.12.2}\cite{Staub:2013tta}. 
1-loop corrected mass spectra and decay branching ratios were calculated with the corresponding \spheno\cite{Porod:2003um,Porod:2011nf} modules, automatically produced by \texttt{SARAH}. 
Low-energy observables, including \deltagmtwomu, have been calculated with \spheno, while to derive the Wilson coefficients $C_{9,\textrm{NP}}^{\mu}$ and $C_{10,\textrm{NP}}^{\mu}$, as well as the parameter $R_{BB}$, we used analytic tree-level formulas~(\ref{C9generic}) and (\ref{bsmix}). Dark matter related observables were calculated with \texttt{MicrOMEGAs~v.4.3.1}\cite{Belanger:2013oya}, based on the \chep\cite{Belyaev:2012qa} files generated by \texttt{SARAH}.
To efficiently scan the multidimensional space of input model parameters, we interfaced all the employed numerical packages to 
\texttt{MultiNest~v.3.10}\cite{Feroz:2008xx} driven by a global likelihood function.


\begin{table}[t]
\begin{center}
\makebox[\textwidth][c]{\begin{tabular}{c|c|c|c|c|c|c|c|c}
\hline
\hline
\rule{0pt}{2.6ex}
Parameter & BP1 & BP2 & BP3 & BP4 & BP5 & BP6 & BP7 & BP8\\
\hline
\hline
\rule{0pt}{2.5ex}
$\mzero$         & $2678$  & $2957$  & $1915$ & $1548$    & $3017$ & $2278$ & $2481$ & $2291$\\
$m_{\tilde{q},33}\!=\!m_{\tilde{u},33}$ & $7156$   & $ 5007$  & $4956$ & $4062$ & $5489$ & $5236$ & $6068$  & $5390$\\
$M_{1,2}$     & $3920$ & $2726$   &  $1913$     & $2463$     & $1906$ & $1839$ & $2924$  & $2128$ \\
$M_{3}$      & $2615$   & $3167$     & $2911$       & $3148$   & $2990$ & $3024$ & $2849$  & $3106$ \\
$\azero$      & $-\,5625$ & $-\,2588$   & $-\,4286$ & $-\,4208$ & $-\,5468$ & $-\,4355$& $-\,2117$ & $-\,5327$ \\
$B_0$          & $-\,918$  & $-\,1486$     & $-\,1456$ & $-\,1159$ & $-\,756$ & $-\,745$    & $-\,1623$ & $-\,1344$\\
$\mu$              & $1013$  & $1143$     & $1509$      & $1128$ & $1265$      & $1725$     & $1044$ & $1098$ \\
$B_{\mu}\,[10^5]$ & $4.8$ & $1.9$   & $5.1$       & $3.5$ & $4.7$             & $2.8$          & $1.4$ & $3.2$ \\
$\tan\beta$          & $47.9$  & $53.3$   & $48.9$ & $47.0$ & $42.3$        & $36.6$     & $35.3$ & $43.0$ \\
\hline
\rule{0pt}{2.5ex}
$M_Q$                 & $1900$  & $2000$       & $2672$ & $2182$ & $2000$  & $2110$ & $1800$ & $1888$  \\
$M_{L}$            & $2230$    & $954$            & $915$ & $2450$ & $2972$     & $1364$ & $815$      & $1174$ \\
$M_{E}$                 & $701$    & $700$        &  $1046$ & $2141$ & $497$        & $1534$ & $2262$    & $801$ \\
$\lam_{Q,2}=\lam_{Q,3}$ & $0.08$  & $0.07$  & $0.80$ & $0.15$ & $0.08$        & $0.80$ & $1.10$       & $0.07$ \\
$\lam_{L,2}$ & $-\,1.2$           & $-0.895$  & $-\,0.579$ & $-\,0.404$& $-\,1.069$ & $-\,1.380$& $-\,1.05$ & $0.436$  \\
$\lam_{E,2}$ & $-\,0.8$        & $0.908$  & $-\,0.746$ & $-\,0.921$ & $-\,0.520$ & $0.705$ & $-\,0.713$ & $-\,1.452$ \\
$\widetilde{Y}_1$ & $0.855$   & $-0.426$  & $0.098$ & $0.450$ & $-\,0.338$ & $-\,0.567$ & $-\,0.615$ & $-\,1.144$  \\
$\widetilde{Y}_2$ & $0.525$  & $0.651$   &  $0.245$ & $-\,0.903$& $-\,0.699$ & $0.388$& $-\,0.958$  & $-\,0.435$ \\
$\mu_S$                & $-\,1270$ & $-\,2161$& $-\,388$  & $968$ & $1125$      & $-\,1764$ & $-\,399$ & $-\,994$ \\
$B_{\mu_S}\,[10^6]$ & $-\,3.5$  & $-\,0.9$ & $-\,1.2$ & $-\,3.9$ & $-\,4.7$ & $-\,3.4$ & $-\,1.3$& $-\,1.0$ \\
$\tan\beta_S$        & $13.6$  & $4.9$      & $0.11$    & $0.73$ & $6.4$     & $0.11$ & $0.06$ & $12.7$ \\
$m_{Z'}$            & $500$ & $351$      &   $310$  &  $472$ & $419$       & $798$ & $430$ & $690$  \\
$M_{\tilde{Z}'}$ & $1943$ & $1592$   & $1151$ & $1385$ & $1844$    & $641$  & $815$ & $491$ \\
$g_X$               & $0.79$& $0.376$ & $0.53$    & $0.66$& $0.62$     & $0.87$ & $0.63$ & $0.76$ \\
$Y_{e,22}$     & $0.016$ & $-\,0.087$  & $0.029$ & $0.026$& $0.034$ & $-\,0.015$ & $-\,0.018$ & $-\,0.012$ \\
\hline
\hline
\end{tabular}}
\caption{Input parameters at the SUSY scale for our benchmark scenarios. Dimensionful quantities are given in~GeV and $\gev^2$.}
\label{tab:benchm_in} 
\end{center}
\end{table}


We show in \reftables{tab:benchm_in}{tab:benchm_sp}eight chosen benchmarks scenarios. 
In \reftable{tab:benchm_in} we present their input parameter values. The top panel of the table gives typical values 
for the relevant MSSM input parameters; the bottom panel holds instead the values associated with the U(1)$_X$ sector.


\begin{table}[t]
\begin{center}
	\makebox[\textwidth][c]{\begin{tabular}{c|c|c|c|c|c|c|c|c}
\hline
\hline
\rule{0pt}{2.5ex}
 & BP1 & BP2 & BP3 & BP4 & BP5 & BP6 & BP7&BP8\\
\hline
\hline
\rule{0pt}{2.5ex}
$m_{\neutone}$ & $1042$  & $1165$  & $328$ & $848$ & $1071$ & $429$ & $257$ & $273$ \\
$m_{\neuttwo}$ & $1044$  & $1168$   & $417$ & $1046$ & $1186$ & $1719$ &$470$ & $1118$ \\
$m_{\charone}$ & $1043$  & $1167$  & $1529$ & $1153$ & $1289$ & $1731$ & $1068$ & $1120$  \\
$m_{\tilde{g}}$ & $2867$   & $3373$  & $3029$ & $3176$ & $3238$ & $3171$ & $3034$ & $3226$\\
$m_h$           & $125.5$  & $123.6$ & $123.2$& $125.6$& $126.0$& $123.6$& $125.1$& $124.6$  \\
$m_{A,H}$       & $3830$  & $2128$  & $4189$ & $3108$ & $3752$ & $1996$ & $1753$ & $2765$  \\
$m_{h'}$        & $487$  & $360$  & $291$  & $423$  & $402$  & $739$  & $406$   & $645$  \\
$m_{A',H'}$     & $6788$  & $2475$  & $3026$ & $5031$ & $5558$ & $5083$ & $4309$ & $3508$ \\
$m_{e_4}$        & $706$  & $706$    & $937$  & $2109$  & $503$ & $1386$  & $830$  & $797$   \\
$m_{e_5}$         & $2301$  & $1139$   & $1094$ & $2490$ & $3026$ & $1602$ & $2256$ & $1226$  \\
$\msusy$           & $3117$  & $3446$  & $2785$ & $2334$ & $3417$ & $2863$ & $3030$ & $2837$  \\
$m_{\tilde{e}_1}$   & $2521$  & $2693$   & $1596$ & $1524$ & $2928$ & $2096$ & $2444$ & $2031$\\
$m_{\tilde{e}_2}$     & $2656$   & $2780$  & $1811$ & $1555$ & $2988$ & $2223$ & $2478$ & $2089$ \\
$m_{\zp}$               & $491$   &   $348$   & $303$ & $465$   & $412$ & $791$  & $423$  & $684$\\
$m_{u_4}\!\approx\!m_{d_4}$& $2164$   & $2227$  & $2436$ & $2433$ & $2246$ & $2379$ & $2035$ & $2140$ \\
\hline
\rule{0pt}{2.7ex}
\abundchi\                              & $0.11$ & $0.12$ \  & $0.10$    & $0.11$ & $0.11$ & $0.10$& $0.11$ & $0.10$ \\
$N_{\textrm{higgsino}}$    				& $99.9\%$& $99.8\%$& $0$       & $0$     & $0$   & $0$   & $0$   & $0$ \\
$N_{\tilde{S}_1}$       				 &$0$          &  $0$     & $53\%$  & $32\%$    & $41\%$ & $18\%$& $57\%$ & $33\%$\\
$N_{\tilde{S}_2}$        				 &$0$   	 &  $0$       &  $40\%$  & $48\%$    & $50\%$ & $2\%$ & $25\%$ & $4\%$\\
$N_{\tilde{Z}'}$  					 &$0$  		 &   $0$   &   $7\%$ & $20\%$    & $9\%$ & $80\%$ & $18\%$ & $63\%$ \\
$\sigsip\times10^{46}$          & $0.63$	&$2.4$ &$1\!\times\!10^{-3}$ &$8\!\times\! 10^{-3}$& $4\!\times\! 10^{-2}$ & $4\!\times\! 10^{-4}$ & $1\!\times\! 10^{-2}$    & $2\!\times\! 10^{-3}$ \\
$\sigma v\times 10^{26}$       & $0.9$        &   $1.0$      &   $1.3$       &    $   2.2  $    &   $2.1$          & $7\!\times\! 10^{-3}$ & $3\!\times\! 10^{-4}$ & $4\!\times\! 10^{-5}$\\
final state                                &  $WW/ZZ$ &$WW/ZZ$&$Z'/h'\, Z'$&    $Z'/h'\, Z'$   &     $Z'/h'\, Z'$  & $\bar{e}_3\,e_3$& $\bar{u}_3\,u_3$& $\bar{u}_3\,u_3$ \\
\hline
\rule{0pt}{2.5ex}
$\deltagmtwomu \!\times\! 10^{9}$ & $2.1$  & $1.6$    & $1.6$         & $1.3$    & $1.4$     & $1.1$              & $1.3$          & $2.3$ \\
$C^{\mu}_{9,\textrm{NP}}$      & $-\,1.04$	& $-\,0.61$& $-\,0.67$ & $-\,0.99$ & $-\,0.93$& $-\,1.02$ & $-\,0.71$    & $-\,0.81$ \\
$C^{\mu}_{10,\textrm{NP}}$   	& $-\,0.03$   &$-\,0.08$ & $0.02$   & $0.01$   & $-\,0.01$ & $0.04$   & $0.01$    & $-\,0.02$ \\
$R_{BB}\times 10^{3}$                   & $2$        & $2$      &  $0.6$     & $2$      & $2$        & $4$            & $0.9$       & $2$  \\
$\textrm{BR}_{h\to\mu\mu}\times10^{4}$& $2.0$ & $4.8$ & $3.5$  & $3.1$      & $4.1$     & $<10^{-4}$ & $3.2$ & $<10^{-4}$ \\
\hline
\rule{0pt}{2.5ex}
\zp\ decays & \multicolumn{8}{c}{$\bar{e},\bar{\nu}_{2,3}\,e,\nu_{2,3}$}   \\
\hline
\rule{0pt}{2.5ex}
\hp\ decays & \multicolumn{5}{c|}{$ZZ$, $WW$} & $\neutone\neutone$& $ ZZ, \! WW$ & $\neutone\neutone$ \\
\hline
\rule{0pt}{2.5ex}
$e_4$ decays & \multicolumn{8}{c}{$e_2\,h'$, $e_2\,Z'$} \\
\hline
\hline
\end{tabular}}
\caption{Spectrum and various observables for our benchmark points. Masses are given in~GeV, spin-independent cross section \sigsip\ in cm${}^{2}$, and present-day annihilation cross section $\sigma v$ 
in cm${}^{3}$/s. We define, as usual, $\msusy=(m_{\tilde{t}_1}m_{\tilde{t}_2})^{1/2}$ as the geometrical average of stop masses.}
\label{tab:benchm_sp} 
\end{center}
\end{table}

In \reftable{tab:benchm_sp} we present physical masses and some mixing parameters of the selected benchmark points, and give the value of the observables 
of interest. All the benchmark points featured in the tables are characterized by \deltagmtwomu\ in agreement with the experimental determination within $2\,\sigma$,
the loop-corrected muon mass within 30\% of the measured value, and  $C_{9,\textrm{NP}}^{\mu}$ and $C_{10,\textrm{NP}}^{\mu}$ in agreement with the estimates of the most recent 
global fits, Eqs.~(\ref{Wilsonfit9}) and (\ref{Wilsonfit10}).

 \subsection{Dark matter\label{sec:BP_DM}}
   
  Depending on the properties of the lightest SUSY particle (LSP), our neutralino DM can be divided into three different categories. 
  Note, however, that we focus here on the states typical of our $Z'$-extended model and refrain from addressing those solutions giving the relic density and \gmtwo\  
  in the MSSM, which generally involve a bino-like neutralino LSP co-annihilating in the early Universe with a slepton, 
or undergoing $s$-channel resonant annihilation mediated by one of the MSSM Higgs fields. (Note that slepton-coannihilation and $A/H$-funnel 
are also the favorite DM mechanisms in minimal MSSM extensions by a pair of VL fermion multiplets of the SU(5) group, 
which also contribute positively to the \gmtwo\ calculation\cite{Choudhury:2017fuu}.)
   \bigskip 

   \noindent $\bullet$ \textbf{BP1-BP2: 1~TeV higgsino and the XENON-1T excess.} 
   Benchmarks BP1 and BP2 feature a $\sim 1\tev$ higgsino LSP ($\mu\approx 1\tev\approx m_{\chi_1^0}\approx m_{\chi_1^{\pm}}\approx m_{\chi_2^0}$), 
which is a well studied DM particle of the MSSM\cite{Profumo:2004at}. Unlike in the MSSM, however, where typically 
 there does not exist a solution for the \gmtwo\ anomaly in the parameter space with a $\sim 1\tev$ higgsino, due to the associated large mass for the sleptons in the spectrum, 
in this model $\deltagmtwomu^{\textrm{BP1,BP2}}> 10^{-9}$.

   The enhanced \gmtwo\ value is due to a combination of the mechanisms described in \refsec{sec:gmtwo}. In 
   the particular case of BP1, the largest contribution is given by new neutralino/slepton loops, which boost the value of the anomalous magnetic moment 
   even in the presence of a fairly large slepton mass,
   thanks to the presence of several neutralino states not far above the $\sim1\tev$ scale, as illustrated in \reffig{fig:g2scalar}(b).
   In the case of BP2
   the presence of a fairly light, $\lesssim 500\gev$ scalar, $h'$, associated with the breaking of the U(1)$_X$ symmetry, 
   introduces a sizable contribution to the scalar/VL lepton loop, as described in \refsec{sec:gmtwo} and illustrated in \reffig{fig:g2scalar}(a).
   Note that in general, these new contributions require sizable, but not abnormally large,  
   new Yukawa couplings: $|\lam_{L,2}|,\,|\lam_{E,2}|\gsim 0.5$.
   
   Interestingly, benchmark points like BP1 and BP2 feature a typical value of the spin-independent cross section, \sigsip, 
   large enough to provide a possible explanation for the slight excess of events observed in the XENON-1T 1-year data\cite{Aprile:2018dbl}. The XENON Collaboration has 
recorded a few events over the background consistent at $\sim 1 \sigma$ with a best-fit point featuring $\sigsip= 4.7\times 10^{-47}\,\textrm{cm}^2$ for a WIMP mass $m_{\chi}=200\gev$, 
or $\sigsip\approx 1 \!-\! 2\times 10^{-46}\,\textrm{cm}^2$ for $m_{\chi}\approx 1\tev$. A spin independent cross section of this size corresponds to a typical gaugino mass of about $2.5-3\tev$.
  For recent reviews of the typical direct and indirect DM detection signatures of the $\sim 1\tev$ higgsino in the MSSM, see\cite{Roszkowski:2017nbc,Kowalska:2018toh}.

\begin{figure}[t]
   	\centering
   		\includegraphics[width=0.28\textwidth]{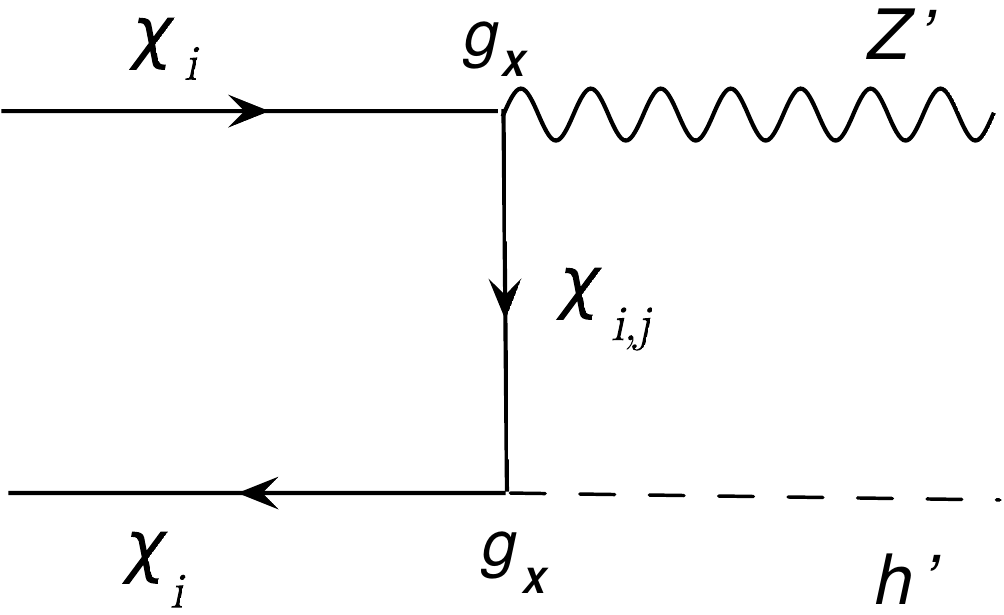}
   	\hspace{0.02\textwidth}
   		\includegraphics[width=0.28\textwidth]{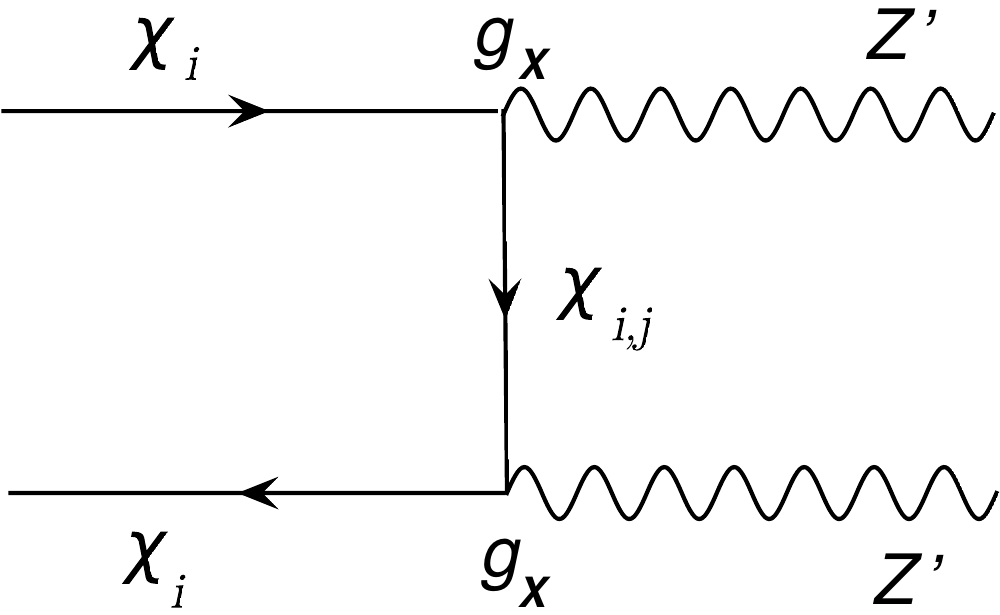}
   	\hspace{0.02\textwidth}
   		\includegraphics[width=0.32\textwidth]{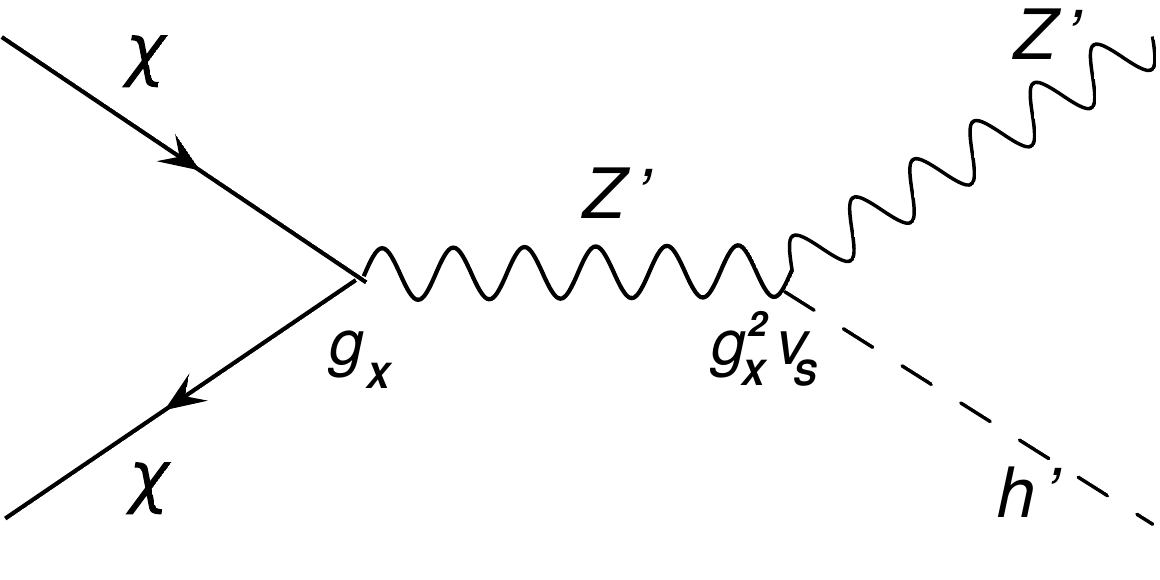}
   	\caption{Example of early-Universe annihilation diagrams for highly-mixed $Z'$-ino/singlino neutralinos, yielding to final states dominated by the $h'$ and $Z'$ bosons.}
   	\label{fig:dm24}
   \end{figure}

   \medskip
   \noindent $\bullet$ \textbf{BP3-BP5: Well-tempered $\boldsymbol{Z'}$-ino/singlino.} The three benchmark points BP3-BP5 represent scenarios in which the neutralino LSP  
   is predominantly singlino, with a non-negligible admixture of $Z'$-ino. The relic density is obtained thanks to a combination of different annihilation diagrams and final states,
   the most dominant of which is a $t$-channel exchange of the second singlino to pair-produce the lighter $Z'$ or $h'$ boson, as depicted in \reffig{fig:dm24}.
   This scenario is reminiscent of the ``well-tempered'' neutralino of the MSSM\cite{ArkaniHamed:2006mb} but, unlike the latter, it involves exclusively the neutralinos of the $Z'$-related sector.  
   
   The parameter space spans over a wide range of neutralino mass values, which can be as heavy as $\sim1\tev$, 
  and of spin-independent cross sections, typically in the range
   $\sigsip= 10^{-50}\!-\!10^{-46}\,\textrm{cm}^2$. Since the dominant interaction of the neutralino with nuclei in underground detectors 
stems in this case from a $t$-channel exchange of the scalar $h'$, its strength is highly dependent on the size of the effective loop couplings that, on the one hand, connect $h'$ to the 
   gluons in the proton, and on the other, generate the mixing of $h'$ with the SM Higgs (recall that at the tree level the scalar mass matrix is block-diagonal).    
   Since to efficiently control the relative size of these effects one would need a detailed analysis of the loop interactions of the model, we do not attempt in this work to connect 
this type of mixed neutralino scenarios to the XENON-1T excess. 

   As pertains to the \gmtwo\ anomaly, the dominant contribution for BP3-BP5 is given by the neutralino/slepton loops, enhanced by the presence of several neutralinos not 
   far above the $\sim 1\tev$ scale, and sizable Yukawa couplings $|\lam_{L,2}|,\,|\lam_{E,2}|\gsim 0.5$.
   
   \medskip
   \noindent $\bullet$ \textbf{BP6-BP8: $Z'$ resonance.} 
   
\begin{figure}[t]
	\centering
	\includegraphics[width=0.33\textwidth]{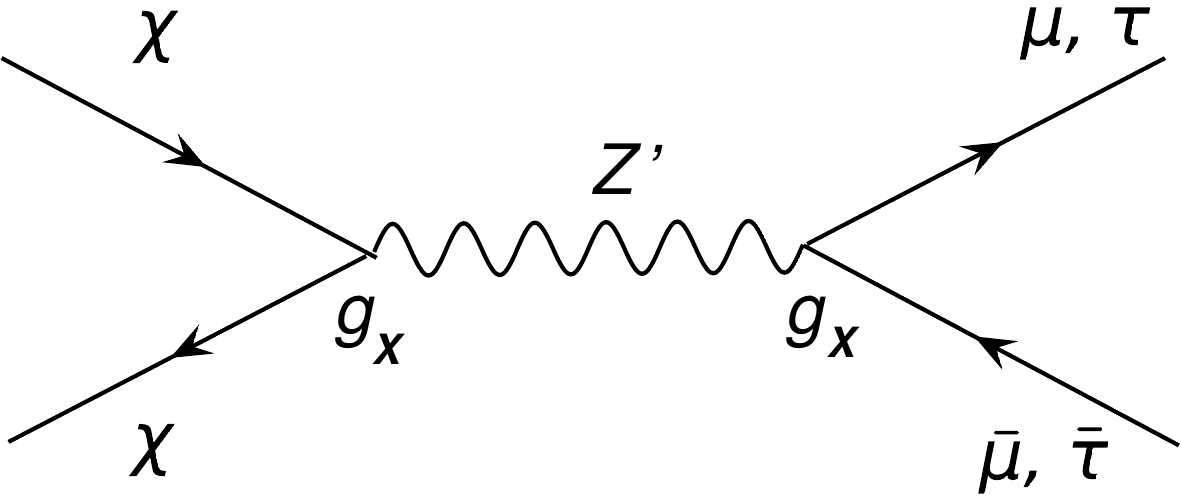}
	\hspace{0.02\textwidth}
	\caption{The dominant early-Universe annihilation channel for a predominantly $Z'$-ino LSP.}
	\label{fig:dm57}
\end{figure}

The last set of benchmark points give examples of neutralinos that show a sizable $Z'$-ino composition, with subdominant or equivalent singlino components. In the early Universe 
an annihilation cross section of the right size for the relic abundance can be obtained through an $s$-channel resonance with the fairly broad $Z'$ boson, yielding to 
final states highly dominated by muons and taus, as is typical of $L_{\mu}-L_{\tau}$
models (see, e.g.,\cite{Sierra:2015fma,Baek:2017sew}). We show the diagram for this process in \reffig{fig:dm57}.

As before, the spin-independent scattering cross section with protons is due to the exchange of a $h'$ boson, and is consequently suppressed by loop effects. At the same time, the present-day 
annihilation cross-section, $\sigma v$, is suppressed by $p$-wave and thermal broadening, so that this scenario is difficult to detect by direct and indirect searches for DM. 
As we shall see in \refsec{sec:exp}, since BP6-BP8 are characterized by a fairly light $Z'$ boson, in the cases where $\tanb_S>1$ the best prospects for detection might be provided by the 
high-luminosity LHC (HL-LHC).

   \begin{figure}[t]
   	\centering
   	\includegraphics[width=0.55\textwidth]{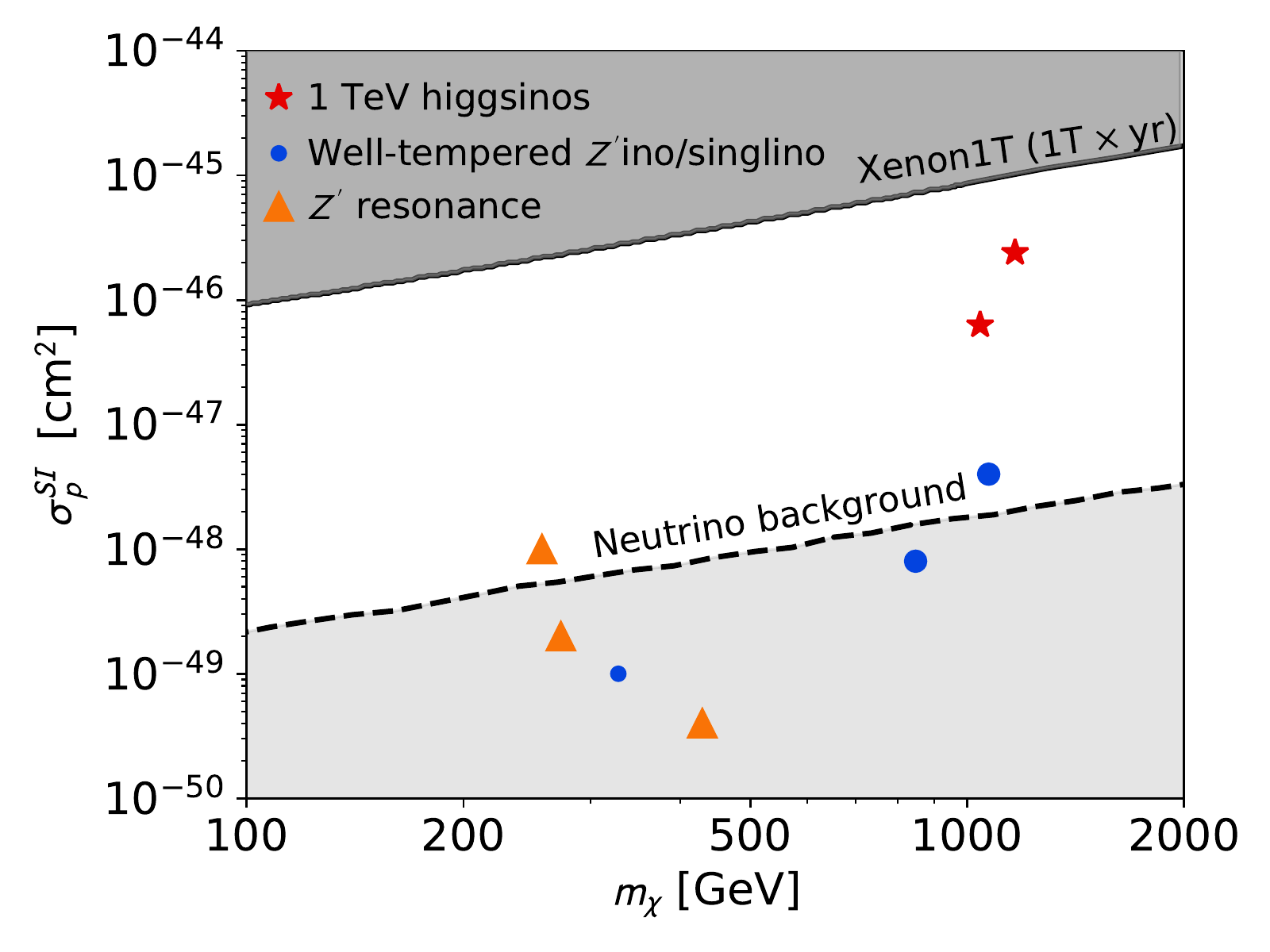}
   	\hspace{0.02\textwidth}
   	\caption{The position of the benchmark points in the ($m_{\chi}$, \sigsip) plane, compared to the current 90\%~C.L. upper bound from the XENON-1T Collaboration\cite{Aprile:2018dbl}. 
Different point-styles mark the different types 
of neutralinos described in \refsec{sec:BP_DM}.}
   	\label{fig:BPsigsip}
   \end{figure}

We summarize in \reffig{fig:BPsigsip} the position of our eight benchmark points in the ($m_{\chi}$, \sigsip) plane. 
The different types of neutralinos described above are marked with different style, according to the legend.

\subsection{Additional constraints on the flavor structure \label{sec:exp}}

As we have seen in the previous sections, a proper fit to the \gmtwo\ and flavor anomalies requires the presence of an extra flavor structure beyond the MSSM. 
While flavor physics (and in particular $B_s$-mixing) provides the dominant constraint on our model, other BSM searches can also be relevant in certain regions of the parameter space. 

\subsubsection*{Neutrino trident production}

Since the $L_{\mu}-L_{\tau}$ gauge boson couples at the tree-level to muon and tau neutrinos, one can expect a strong enhancement in the neutrino trident production from scattering on atomic nuclei $N$:
$N \nu \rightarrow \nu N \mu^+ \mu^-$\cite{Altmannshofer:2014pba}. 
The cross section for this process has been measured by the CCFR\cite{Mishra:1991bv} and CHARM-II\cite{Geiregat:1990gz} collaborations to be in agreement with the SM prediction, 
leading in the large $m_{\zp}$ limit ($m_{\zp}\gsim 10\gev$) to a generic bound:
\begin{align}
m_{\zp}/g_X \gtrsim  540 \text{ GeV}\ .
\end{align}
This constraint typically provides a lower bound on the accessible $m_{\zp}$ mass, independent of the remaining parameters of the flavor sector.

\subsubsection*{LHC Higgs measurements in the muon channel}

We have mentioned in \refsec{sec:gmtwo} that, since the largest contributions to the calculation of \gmtwo\ involve loop diagrams that violate chiral symmetry, 
one must proceed to tune the Yukawa coupling of the muon against large loop corrections to the muon mass.
Thus, a powerful constraint on the benchmark points can be derived from measurement of SM Higgs decay in the $\mu\mu$ channel, which test directly the effective muon Yukawa coupling. The current experimental signal strength for this channel is $\sigma/\sigma^{SM} = 0 \pm 1.3$\cite{Tanabashi:2018}. 

Since the mixing between the new Higgs sector and the MSSM one is loop-suppressed, the production mechanism for the 125\gev\ Higgs boson is to a good approximation similar to the one of the SM. Hence the quantity that determines directly the signal strength is the branching ratio to muons, 
which in the SM reads $\textrm{BR}(h_{\textrm{SM}}\rightarrow \mu\mu)=2.2\times 10^{-4}$\cite{Tanabashi:2018}.  We report the corresponding value of the branching ratio for our benchmark points at the bottom of \reftable{tab:benchm_sp}. Note that in all benchmark points we are within the
current upper bound $\textrm{BR}(h_{\textrm{SM}} \rightarrow \mu\mu) < 5.7\times 10^{-4}$, even if all points are characterized by significant loop contributions to \gmtwo.

At the tree level, the branching ratio is directly proportional to the effective Yukawa coupling squared, $|Y_{\textrm{eff},22}|^2$, 
between the SM Higgs, $h_{\textrm{SM}}$, and the muon:
$\mathcal{L}\supset -Y_{\textrm{eff},22}\,(h_{\textrm{SM}}\,e_{L,2}\,e_{R,2}$ $+$ $\textrm{h.c.})$.
In our model, this effective coupling can be readily expressed as
\bea
Y_{\textrm{eff},22}&=& Y_{e,22}\left(-\sin\alpha\right)D'^{\dag}_{L,22}D'^{\dag}_{R,22}+\widetilde{Y}_1\left(-\sin\alpha\right)D'^{\dag}_{L,24}D'^{\dag}_{R,25}+\widetilde{Y}_2\left(\cos\alpha\right)D'^{\dag}_{L,25}D'^{\dag}_{R,24}\,, \nonumber
\eea
where we use the rotation matrix elements defined in \refeq{diagproc}, and the Higgs rotation angle $\alpha$ is defined in Appendix~\ref{app:model}. In the large $\tan \beta$ limit, we find
\begin{align}\label{effectYuk}
|Y_{\textrm{eff},22}|^2~\approx~ \frac{1}{\tan^2 \beta} \frac{1}{\left(1+\frac{\lam_{E,2}^2 v_2^2}{2 M_E^2}\right)\left(1+\frac{\lam_{L,2}^2 v_1^2}{2 M_L^2}\right)} \left( Y_{e,22} + \widetilde{Y}_1 \frac{v_1 v_2 \lambda_{E,2} \lambda_{L,2}}{2 M_E M_L} \right)^2.
\end{align}
In particular, the effective Yukawa coupling contains a term originating from the new VL sector. Hence, depending on the values of $Y_{e,22}$ and $\widetilde{Y}_1$ the two terms can cancel out, leading to a branching ratio significantly lower than the SM value. This typically occurs in our benchmark points BP6 and BP8.

The main consequence of these effects is that, by increasing the sensitivity of Higgs searches in the muon channels, the LHC collaborations can constrain the Yukawa couplings of our models, which in turn will refine the predictions of our model for \gmtwo\ and flavor observables in general.

\subsubsection*{LHC exotica searches}

The MSSM sector of the model could be in principle tested by a wide set of LHC SUSY searches. 
However, since all the anomalies discussed in \refsec{sec:flav} can be accommodated by setting 
the masses of the MSSM gluinos, charginos, squarks, and sleptons above the corresponding experimental lower bounds, we do not consider the impact of these searches in this work. 

On top of the MSSM particles, the model presented in this paper can feature several new fields in the sub-TeV regime, including the new gauge boson an additional neutral Higgs boson, 
and the fourth and fifth-generation VL leptons. 
In the following, we will study the bounds on these states from existing LHC searches, to ensure that the benchmark points presented in \reftables{tab:benchm_in}{tab:benchm_sp}are compatible with the experimental data. In our numerical analysis we use the UFO model file generated by \texttt{SARAH~v.4.12.2}\cite{Staub:2013tta} and passed to \madgr\cite{Alwall:2014hca} to estimate the relevant LO cross sections times the branching ratio quantities.

\paragraph{ $\boldsymbol{Z^\prime}$ signatures}

The new gauge boson \zp\  couples to the SM fields either through its mixing with the $Z$ boson (which we assume to be negligible in this study), 
or through the mixing between the  U(1)$_X$-charged VL  quarks and their  SM counterparts. 

\begin{figure}[t]
	\centering
	\subfloat[]{%
		\label{fig:a}%
		\includegraphics[width=0.3\textwidth]{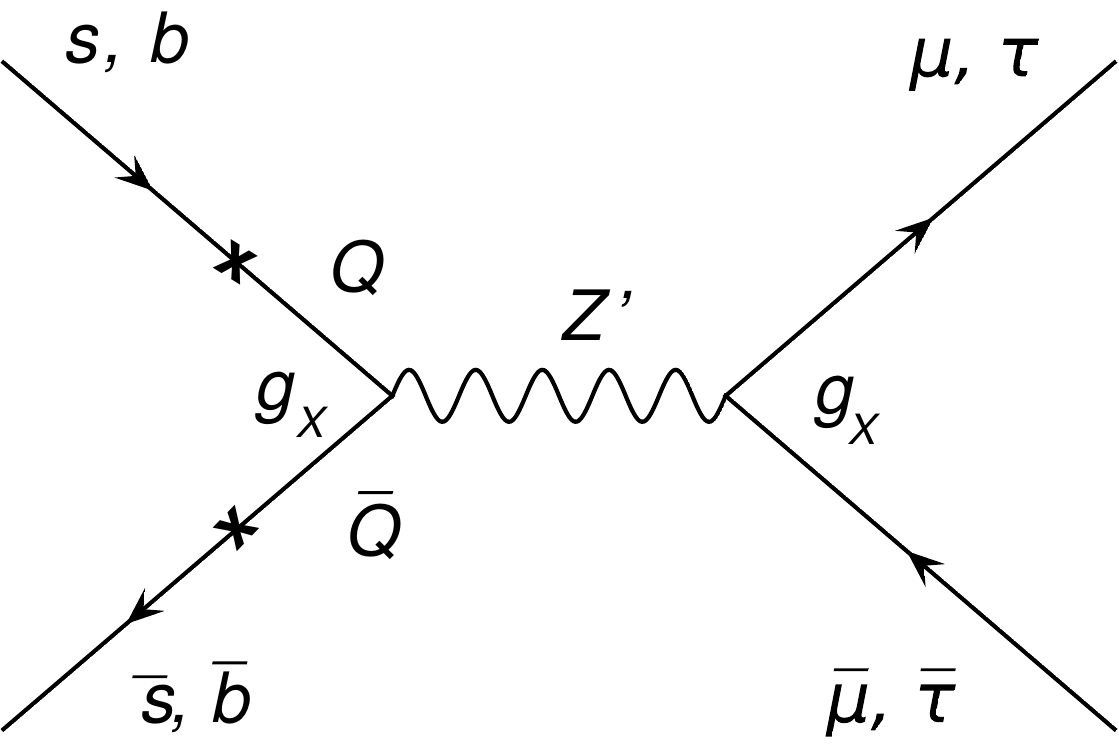}
	}%
	\hspace{0.17\textwidth}
	\subfloat[]{%
		\label{fig:b}%
		\includegraphics[width=0.4\textwidth]{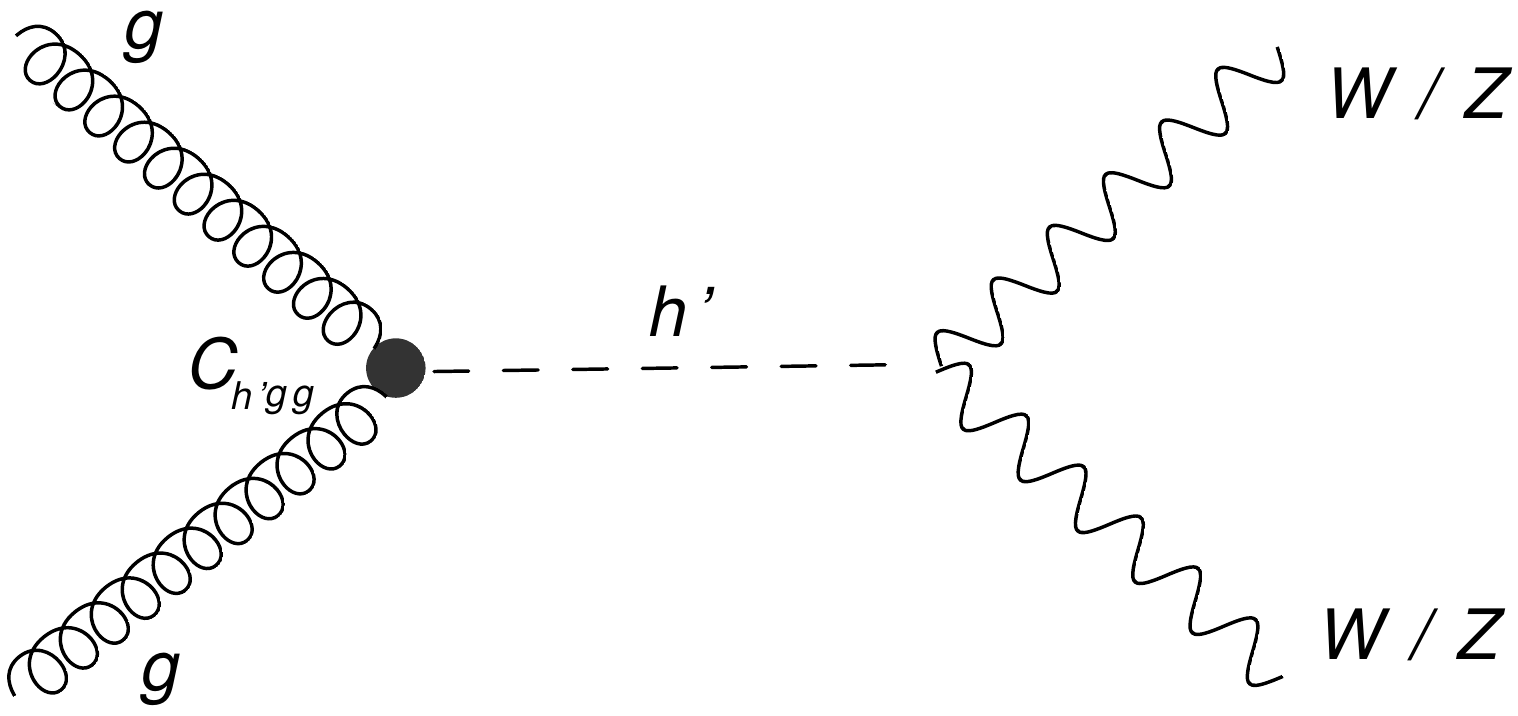}
	}%
	\caption{The dominant production and decay channels for the beyond the MSSM bosons. (a) Heavy quark fusion production of \zp. (b) Gluon fusion production of $h'$. }  
	\label{fig:zhprime}
\end{figure}

The dominant production mechanism of \zp\ is heavy quark fusion, as shown in \reffig{fig:zhprime}(a). Since it can only proceed through the mixing of $Q$ with the $s$ and $b$ quarks, it is PDF-suppressed and crucially depends on the value of $v_{1}$, see \refeq{deltaLbs}. This implies that \zp\ production is strongly suppressed when $\tan \beta_S \ll 1$.\footnote{Note that the assumption about no kinetic mixing is crucial here, as otherwise a constant contribution proportional to the mixing parameter $\epsilon'$ would be generated.} 


\begin{figure}[t]
	\centering
	\subfloat[]{%
		\label{fig:a}%
		\includegraphics[width=0.49\textwidth]{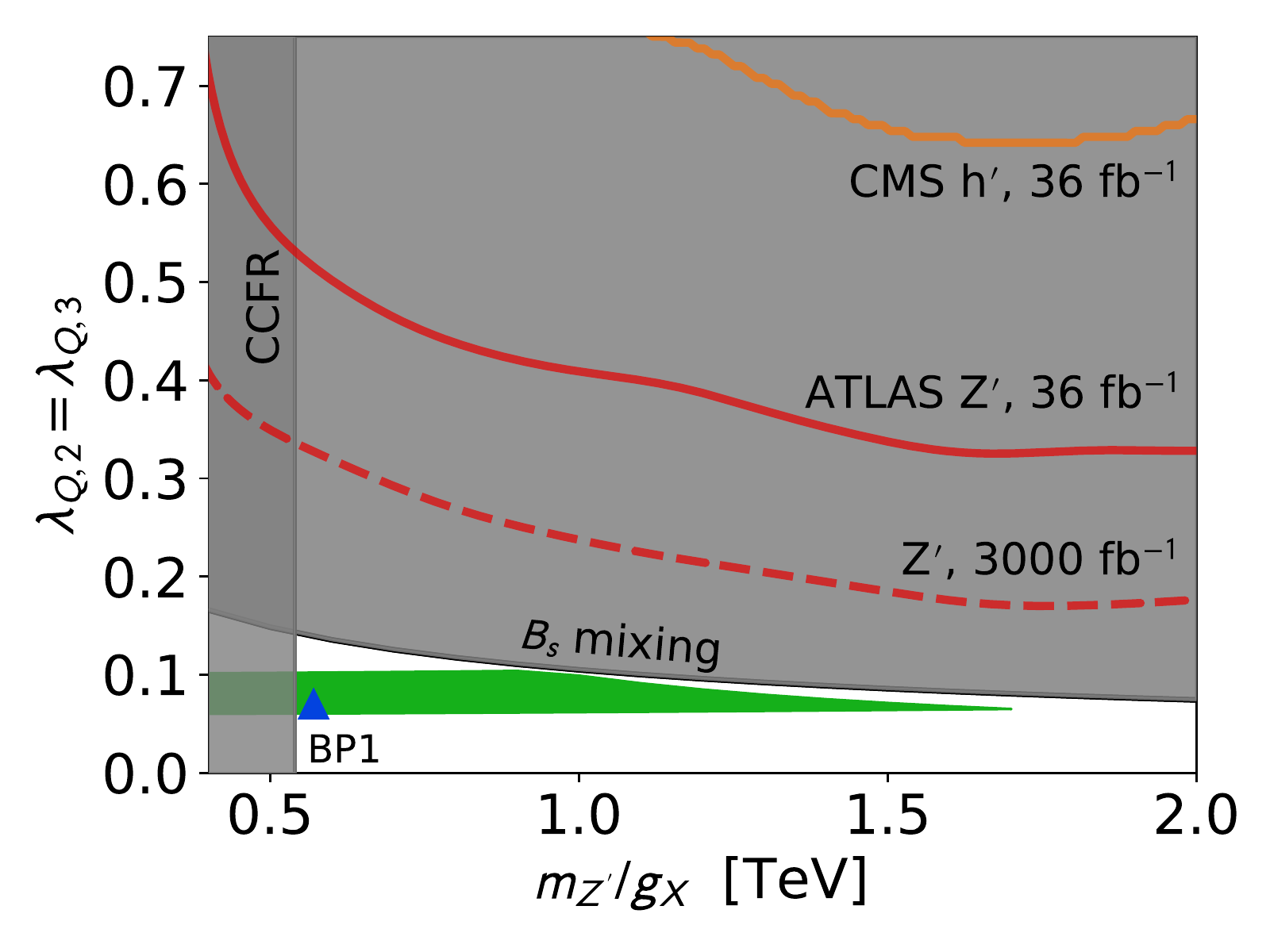}
	}%
	\subfloat[]{%
	\label{fig:b}%
	\includegraphics[width=0.49\textwidth]{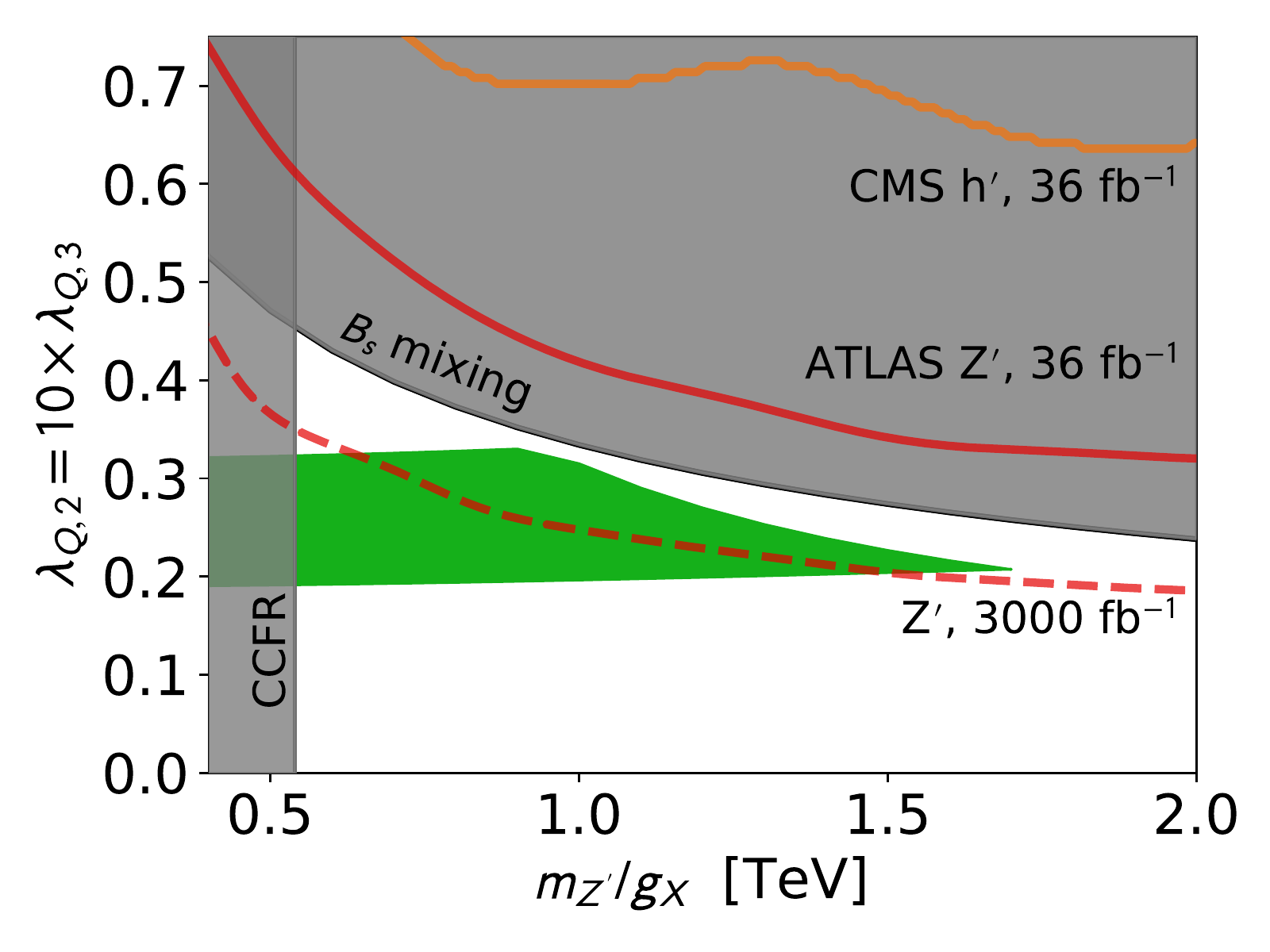}
}%
	\caption{Bounds from $B_s$-mixing (gray area is excluded) and neutrino trident production (dark gray area is excluded). Upper exclusion bound from current ATLAS\cite{Aaboud:2017buh} 
dilepton search (red line), our HL-LHC projection for the same analysis (dashed red line), di-$Z$ search from $h'$ decay (orange line) 
and the region favored by the $R_{K^{(\ast)}}$ anomalies (green area), in the $\lambda_{Q,2}$ and 
		$ m_{\zp} /g_X$ plane for the ratios (a) $\lambda_{Q,2} = \lambda_{Q,3}$, (b) and $\lambda_{Q,2} = 10 \times \lambda_{Q,3}$. All other parameters are chosen according to benchmark point BP1 (blue star) as presented in \reftables{tab:benchm_in}{tab:benchm_sp}.}  
	\label{fig:BP1_RKZp}
\end{figure}


The dominant visible decay channel of \zp\ is into a pair of muons or taus. The strongest experimental limits at $\sqrt{s}=13\tev$ are provided by the ATLAS\cite{Aaboud:2017buh} and CMS\cite{Sirunyan:2018exx} high-mass resonance searches  based on the dilepton channels for 36\invfb\ luminosity. The limits apply to the product of production cross section times the branching fraction into a pair of muons, as a function of the \zp\ mass. In \reffig{fig:BP1_RKZp}(a) we show in red solid the 95\%~C.L. dilepton exclusion upper bound based on the ATLAS analysis as a function of the new Yukawa coupling $\lambda_{Q,2}$ and of the \zp\ mass, for the choice of other model parameters corresponding to BP1 (depicted with a blue triangle). We also present, as a red dashed line, 
our estimate of the potential reach of the HL-LHC with an integrated luminosity of 3000\invfb.

Incidentally, note that the parameters of interest for the \zp\ production are $M_Q$, $\lambda_{Q,2}$, $v_1$ and $g_X$, the same that also enter the $B_s$ mixing. However, in contrast to the upper 
bounds from $B_s$-mixing (depicted in \reffig{fig:BP1_RKZp}(a) as a gray shaded region), the \zp\ production relies mostly on the mixing between the VL quark $Q$ and the strange quark, and hence on the product $v_1 \lambda_{Q,2}$. This implies that when the Yukawa couplings of the VL quarks present an inverted hierarchy, $\lambda_{Q,3} \ll \lambda_{Q,2}$, the exclusion bound bites more deeply into the region of the parameter space favored by the LFUV anomalies, shown here in green. We illustrate this effect in \reffig{fig:BP1_RKZp}(b), in which we assume that $\lambda_{Q,2}=10\,\lambda_{Q,3}$. This feature opens up the possibility of experimentally testing the structure of the BSM flavor sector with the HL-LHC.
Finally, the fact that the \zp\ is dominantly produced through its coupling to strange quarks sets apart the low-energy limit of our model  from the effective approach of\cite{Kohda:2018xbc} where it was found that a discovery at HL-LHC required large $B_s$-mixing values.

We do not show the corresponding exclusion bounds for other benchmark points from \reftable{tab:benchm_in}. As long as $\tan \beta_S \gg 1$ (BP2, BP5, and BP7), their dependence on the parameters of interest is exactly the same as the one shown in \reffig{fig:BP1_RKZp}(a). If, on the other hand, $\tan \beta_S \ll 1$ (as is the case of BP3, BP4, BP6, and BP7), \zp\ production is strongly suppressed at the LHC 
and the dilepton searches do not provide any constraints for perturbative values of $\lambda_{Q,2}$.

\medskip

\paragraph{New light Higgs decays}

In order to open up an efficient mechanism for pair annihilation of the neutralino LSP in the early Universe, for BP3-BP5 the lightest of the two new neutral Higgs bosons, $h'$, has a mass in the sub-TeV regime. Similarly the sub-TeV \zp\ and $h^\prime$ are favored by \gmtwo\ in the case of BP1 and BP2 and by the \zp\ resonance for BP6-BP8.
When  kinetic mixing is negligible,  $h^\prime$ interacts with the SM particles either through its (loop-induced) mixing with the SM Higgs, or through the loops mediated by the new VL fermions. The dominant production channel of $h'$ is gluon fusion, shown in \reffig{fig:zhprime}(b), with an effective coupling to gluons $C_{h'gg}$ generated via the two above-mentioned mechanisms. The former typically dominates for lighter $h'$, while the latter for heavier.

Decays of the new Higgs boson are driven mostly by its mixing with the SM Higgs, with the highest branching ratios into $h'\to W^+W^-$ and $h'\to ZZ$ for the benchmark points shown in  \reftable{tab:benchm_in}.
Such decay channels have been studied by both the ATLAS and CMS collaborations, with the most recent results reported in\cite{Aaboud:2017gsl,Aaboud:2017fgj,Aaboud:2017itg} for the former,  and\cite{Sirunyan:2018qlb} for the latter. While these searches can typically probe $\sigma \times \textrm{BR}$ values down to $10-100$ fb, in our model $h'$ production is typically very suppressed by being loop-induced. Overall, we find that the corresponding experimental  bounds, illustrated in \reffig{fig:BP1_RKZp}\subref{fig:a} as an orange solid line, are weaker than the ones from the $\zp$ searches. 

More generally, we observe that also in this case $\tan \beta_S$ is the key parameter that allows to access the new sector experimentally. Since only the $S_1$ scalar couples directly to the new heavy VL quarks, see \refeq{superpot}, for $\tan \beta_S \ll 1$ the new light Higgs boson is effectively secluded from the SM.

\medskip

\paragraph{Masses of VL fermions}
The new VL leptons can be pair-produced at the LHC via Drell-Yan. Their preferential decay modes are $e_{4,5}\rightarrow e_{2}\,\zp$, $e_{4,5}\rightarrow e_2\,h^\prime$ for lighter $Z'$, 
and $e_{4,5}\rightarrow e_{2}\,Z$, $e_{4,5}\rightarrow e_2\,h$ when the \zp\ is heavier than all of the leptons. 
Multi-lepton searches at the LHC typically set constraints on these second type of decay chains, which translate into a lower bound on the VL lepton mass of the order of $M_L,M_E\gsim200-300\gev$, 
depending on the branching ratio\cite{Dermisek:2014qca,Kumar:2015tna}. 

In the bulk of our parameter space, however, the 4th and 5th lepton generations are generically heavier and decay preferentially via $\zp$ and $h^\prime$. While the former lead to final states similar to the standard $Z$-induced decay (but with different kinematics due to the heavy mass of the $\zp$), the latter typically implies final states with larger lepton and/or jets multiplicities, as $h'$ decays in turn into a pair of $W$ or $Z$ bosons. No existing LHC bounds actually apply to these states, and deriving a constraint would require a full recasting, which exceeds the purposes of this work. Let us simply note that the pair production cross section for such states is typically very small: $\sigma_{pp \rightarrow L' L'} \approx 2 \fb$ 
for VL lepton masses of around $500\gev$.

It is easier to pair-produce the new VL quarks. Searches for heavy tops/bottoms have been regularly updated by both the ATLAS and CMS collaborations, 
with exclusion bounds implying $M_Q \gsim 900-1300\gev$ depending on the branching ratio of the VL particles. These searches generally assume decay channels to $b/t$ mediated by $h/W/Z$. 
In our model $Q$ and $Q^\prime$ often dominantly decay emitting an $h'$ or $\zp$ instead of a SM boson.
As $\textrm{BR}(Z'\rightarrow \nu_{\mu,\tau}\,\nu_{\mu,\tau})\approx 30\%$, the most relevant (and directly applicable) search is the ATLAS Collaboration bound\cite{Aaboud:2017qpr}, on $Q\rightarrow tZ$ followed by $Z \rightarrow \nu \nu$.
It reads $M_Q\gsim 1\tev$. Interestingly, the $L_{\mu}-L_{\tau}$ nature of the $\zp$ implies that it decays almost exclusively into leptons, making VL searches based on multi-jets final state less-sensitive. 

Overall, since the flavor anomalies considered in this study only loosely constrain the mass of VL quarks, we choose $M_Q$ in the multi-TeV region, safely away from all current LHC bounds.

\subsubsection*{Electroweak precision}

Another class of low-energy observables that can be affected by BSM physics above the EWSB scale are the electroweak precision observables (EWPOs). Within the model considered in this study they can in principle originate from different sources. SUSY is known to not generate large contributions to EWPO constraints\cite{Cho:1999km} as it decouples quickly with \msusy. Extra scalar multiplets could in principle be responsible for measurable deviations from the experimental data if their components were not mass-degenerate. This is, however, not the case in this study since $S_1$ and $S_2$ are SM singlets. Therefore, the only source of extra contributions to EWPOs are VL fermions $Q,Q'$, $L,L'$, and $E,E'$. Since the VL fermions couple to the gauge bosons with vector couplings only, and we assume in this study that the kinetic mixing between $Z$ and \zp\ is negligible, their contribution to EWPOs is expected to be proportional exclusively to the mixing with the chiral fermions of the SM.

The first class of BSM corrections  we analyze are the {\it oblique} corrections to the gauge bosons' vacuum polarization, quantified by the Peskin-Takeuchi parameters $S$, $T$ and $U$\cite{Peskin:1991sw}. The size of the parameter $T$ depends on the difference between the $Z$ and $W$ bosons' self-energies, and as such is sensitive to the mass splitting between the components of electroweak doublets\cite{Olive:2016xmw},
\be
T=\frac{1}{32\pi^2 v^2\alpha}\sum \Delta m^2,
\ee
where $\alpha$ is the fine-structure constant, and 
\be
\Delta m^2=m_1^2+m_2^2-\frac{2m_1^2m_2^2}{m_1^2-m_2^2}\ln\frac{m_1^2}{m_2^2},
\ee 
where we denoted as $m_1$ and $m_2$  the masses of the doublet's components.
In our scenario both components of the VL doublets $Q,Q'$ and $L,L'$ remain degenerate at the tree level, so that the resulting  contribution to $T$ is very small.

The oblique parameter $S$ is defined as\cite{Peskin:1991sw}
\be
\alpha\,S=\frac{\sin^2(2\theta_W)}{m_Z^2}\left[\Pi_{ZZ}(m_Z^2)-\Pi_{ZZ}(0)-\Pi_{\gamma\gamma}(m_Z^2)-2\cot(2\theta_W)\Pi_{\gamma Z}(m_Z^2)\right]\,.
\ee
Vacuum polarizations scale like $\Pi_{AA}\sim c_{L,R}^2$, where $c_L$ and $c_R$ denote couplings of the left- and right-handed fermions to the corresponding gauge bosons, in analogy to the ones introduced in \refsec{sec:gmtwo}. In our scenario the size of the effect is directly proportional to the mixing of the VL fermions with the ones of the SM,
\bea
c_L^{d_i}\sim c_L^{\textrm{SM}}(D_{L\,i 2})^2,&\qquad c_R^{d_i}\sim c_R^{\textrm{SM}}(D_{R\,i 2})^2\,,&\qquad i=4\,,\nonumber\\
c_L^{e_j}\sim c_L^{\textrm{SM}}(D'_{L\,j 2})^2,&\qquad c_R^{e_j}\sim c_R^{\textrm{SM}}(D'_{R\,j 2})^2\,,&\qquad j=4,5\,,
\eea
where $c_{L,R}^{\textrm{SM}}$ denote the corresponding left- and right-handed couplings of the SM chiral fermions. As a result, the BSM contributions to $S$ are strongly suppressed by a small mixing of the BSM sector with muons and quarks $b$ and $s$. The corresponding suppression factors in the lepton sector for the benchmark points are presented in \reftable{tab:ewpo}. The corresponding suppressions in the quark sector are of the order of $10^{-4}$.


\begin{table}[t]
	\begin{center}
		\begin{tabular}{c|c|c|c|c|c|c|c|c}
			\hline
			\hline
			\rule{0pt}{2.5ex}
			& BP1 & BP2 & BP3 & BP4 & BP5 & BP6 & BP7 & BP8 \\
			\hline
			\hline
			\rule{0pt}{2.5ex}
			$c_L^{e_4}/c_L^{\textrm{SM}}$ & $10^{-6}$ & $10^{-3}$ & $10^{-4}$ & $10^{-3}$ & $10^{-6}$ & $10^{-3}$ & $10^{-3}$ & $10^{-4}$\\
			$c_L^{e_5}/c_L^{\textrm{SM}}$ & $10^{-2}$ & $10^{-2}$ & $10^{-5}$ & $10^{-2}$ & $10^{-2}$ & $10^{-5}$ & $10^{-7}$ & $10^{-2}$\\
			$c_R^{e_4}/c_L^{\textrm{SM}}$ & $10^{-4}$ & $10^{-2}$ & $10^{-3}$ & $10^{-2}$ & $10^{-3}$ & $10^{-4}$ & $10^{-5}$ & $10^{-3}$\\
			$c_R^{e_5}/c_L^{\textrm{SM}}$ & $10^{-8}$ & $10^{-4}$ & $10^{-2}$ & $10^{-4}$ & $10^{-8}$ & $10^{-1}$ & $10^{-2}$ & $10^{-6}$\\
			\hline
			\hline
		\end{tabular}
		\caption{Suppression factors of the VL lepton-gauge boson couplings w.r.t. the SM ones.}
		\label{tab:ewpo} 
	\end{center}
\end{table}

Finally, contributions from VL fermions to the oblique parameter $U$ are generally much smaller and can be neglected.

The {\it non-oblique} corrections, i.e., one-loop corrections to the coupling of the $Z$ boson with the left-handed and right-handed muons can arise due to their mixing with the BSM leptons $L,L'$ and $E,E'$.
Again, since they are given by
\be
\Delta g_{L,R}^{\mu}\sim \sum_{j}(c_{L,R}^{e_j})^2\,,
\ee 
the extra sector contribution is strongly suppressed.

\section{Summary and conclusions}
\label{sec:sum}

In this work we considered a supersymmetric version of a model with an $L_\mu - L_\tau$ gauge symmetry and 
the addition of VL fermions. In several studies SM extensions based on U(1)$_{L_\mu - L_\tau}$ without SUSY
have been shown to provide arguably the most natural framework for an explanation of the recent LHCb flavor anomalies. 
In this work we showed that, by further introducing supersymmetry, the model becomes endowed with several welcome additional features, on top of reproducing the Wilson coefficients 
$C^\mu_{9,\textrm{NP}}$ and $C^\mu_{10,\textrm{NP}}$ favored by the global effective field theory fits. 

First and foremost, the SUSY model can show excellent agreement with the long-standing experimental anomaly in \gmtwo, 
and naturally provide a WIMP DM candidate, in the form of the lightest neutralino. 
These are not in themselves features exclusive to the
SUSY extension, as some papers have already pointed out that parameter space can be found in non-SUSY U(1)$_X$ models where 
viable solutions for thermal DM and \gmtwo\ coexist.
What is, however, a peculiar feature of the SUSY extension is that the regions of the 
parameter space abiding to \textit{both} the relic density and \gmtwo\ constraints can accommodate a DM particle as heavy as the TeV scale, due to the presence 
of new Higgs fields and additional neutralinos in the spectrum, which show sizable interactions with the muon and charged sleptons.  
This feature places our model in contrast to both its non-SUSY counterparts and the MSSM, where the measurement of the muon anomalous magnetic moment typically
constrains the WIMP to a maximum mass of a few hundred~GeV.

Interestingly, among those TeV-scale WIMPs associated with a solution to the flavor anomalies we find the well-known $\sim1\tev$ 
higgsino, which in the MSSM is the typical DM accompanying sparticles in the multi-TeV range, and thus emerges as favorite in  
global statistical SUSY analyses after the discovery of the Higgs boson at 125\gev.
As the characteristic scattering cross section of the higgsino with nuclei is expected to be within the near-future reach of underground direct detection experiments, 
an exciting possibility opens up, that the model considered here might 
provide an explanation for the very recent slight event excess in the large DM mass region reported by the XENON-1T Collaboration. 

We finally showed that the constraints from flavor precision measurements and direct BSM searches at the LHC do not currently probe the parameter space relevant for the analyzed muon anomalies. 
At the same time we pointed out that some novel experimental signatures can arise, possibly allowing the model to be tested at the LHC in searches for VL fermions. 
More precisely, the VL quarks and leptons mostly decay by emitting a heavy boson $h'$ or $\zp$ instead of a SM one as is usually assumed in standard searches.  
This typically leads to leptonically-rich final states 
and to leptons with higher $p_T$ than in standard VL searches. 
More generally, the fact that the $\zp$ originates from a $L_{\mu}-L_{\tau}$ gauge symmetry implies that observing a signal in multi-lepton VL 
searches, but not in multi-jets, could play the role of smoking gun for this type of models.

\bigskip
\noindent \textbf{Acknowledgments}
\medskip

\noindent
LD and LR are supported in part by the National Science Centre (NCN) research grant No.~2015/18/A/ST2/00748. LR is also supported by the project ``AstroCeNT: Particle Astrophysics Science and Technology Centre" carried out within the International Research Agendas programme of the Foundation for Polish Science co-financed by the European Union under the European Regional Development Fund. 
KK is supported in part by the National Science Centre (Poland) under the research grant No.~2017/26/E/ST2/00470.
EMS is supported in part by the National Science Centre (Poland) under the research grant No.~2017/26/D/ST2/00490. The use of the CIS computer cluster at the National Centre for Nuclear Research in Warsaw is gratefully acknowledged.

\newpage
\appendix

\section*{Appendix}
\addcontentsline{toc}{section}{Appendices}

\section{Some details of the model\label{app:model}}

We provide in this appendix some of the technical details of our model. 
The superfield content and gauge quantum numbers with respect to the four gauge groups are summarized in \reftable{tab:qn}. 

\bigskip
\noindent \textbf{Fermion masses}

The explicit tree-level form of the charged lepton mass matrix reads, in the basis $(e_L,\mu_L,\tau_L,E_L,E')$ $\times$ $(e_R,\mu_R,\tau_R,E_R,E)$,
\be
\mathcal{M}_{e}=\left(\begin{array}{ccccc}
	Y_{e,11} v_d & 0 & 0 & 0 & 0 \\
	0 & Y_{e,22} v_d & 0 & \lam_{L,2} v_1/\sqrt{2} & 0 \\ 
	0 & 0 & Y_{e,33} v_d& \lam_{L,3} v_1/\sqrt{2}& 0 \\
	0 & 0 & 0 & -M_L & \widetilde{Y}_1 v_d  \\
	0 & -\lam_{E,2} v_2/\sqrt{2} & -\lam_{E,3} v_2/\sqrt{2} & \widetilde{Y}_2 v_u & M_E 
\end{array}\right)\,,
\ee
where $v^2\equiv v_u^2+v_d^2=174\gev$ and $v_S^2\equiv v_1^2+v_2^2=m_{Z'}^2/g_X^2$.
We adopt throughout this paper the assumption $\lam_{L,3}\approx\lam_{E,3}\approx 0$.

The explicit form of the tree-level down-type quark mass matrix, in the basis $(d_L,s_L,,b_L,D_L)$ $\times$ $(d_R,s_R,b_R,D_R)$, reads
\be
\mathcal{M}_d=\left(\begin{array}{cccc}
	&  &  & \lam_{Q,1} v_1/\sqrt{2} \\
	& Y_{d,ij}\,v_d &  & \lam_{Q,2} v_1/\sqrt{2} \\ 
	&  &  & \lam_{Q,3} v_1/\sqrt{2} \\
	0 & 0 & 0 & -M_Q 
\end{array}\right)\,,\label{bmatrix}
\ee
where we assume throughout this work that $\lam_{Q,1}\approx 0$.

As was mentioned in \refsec{sec:rk}, we diagonalize the fermion mass matrices by means  
of unitary matrices, $D_L,D_R$ and $D_L',D_R'$, such that 
\be
\mathcal{D}_d=D_L^{\dag} \mathcal{M}_d D_R\,, \quad \mathcal{D}_e=D_L'^{\dag} \mathcal{M}_e D_R',
\ee
where $\mathcal{M}_d$ and $\mathcal{M}_e$ are, respectively, the down-type quark and charged-lepton mass matrices 
presented above and $\mathcal{D}_d$ and $\mathcal{D}_e$ are the corresponding diagonal matrices.
Mass eigenstates $d_i, e_i$ in terms of gauge eigenstates $d'_i, e'_i$, where the former are ordered by increasing mass, and the latter are ordered 
as in the bases given above, are given by 
\be
d_{L(R)\,j}=\sum_r D^{\dag}_{L(R)\,jr}d_{L(R)\,r}'\,, \quad e_{L(R)\,j}=\sum_r D'^{\dag}_{L(R)\,jr}e_{L(R)\,r}'\,.
\ee

\begin{table}[t]
	\begin{center}
		\begin{tabular}{c|c|c|c|c}
			\hline
			\hline
			\rule{0pt}{2.5ex} \small{} Field  &    SU(3)   &  SU(2)$_L$ & U(1)$_Y$ & U(1)$_X$     \\
			\hline
			\hline
			\rule{0pt}{2.5ex} $l_1$ &  $\mathbf{1}$  & $\mathbf{2}$ & $-1/2$ & $0$ \\
			\rule{0pt}{2.5ex} $l_2$ &  $\mathbf{1}$  & $\mathbf{2}$ & $-1/2$ & $+1$ \\
			\rule{0pt}{2.5ex} $l_3$ &  $\mathbf{1}$  & $\mathbf{2}$ & $-1/2$ & $-1$ \\
			\rule{0pt}{2.5ex} $e_R$ &  $\mathbf{1}$  & $\mathbf{1}$ & $+1$ & $0$ \\
			\rule{0pt}{2.5ex} $\mu_R$ &  $\mathbf{1}$  & $\mathbf{1}$ & $+1$ & $-1$ \\
			\rule{0pt}{2.5ex} $\tau_R$ &  $\mathbf{1}$  & $\mathbf{1}$ & $+1$ & $+1$ \\
			\rule{0pt}{2.5ex} $q_{i=1,2,3}$ &  $\mathbf{3}$  & $\mathbf{2}$ & $+1/6$ & $0$ \\
			\rule{0pt}{2.5ex} $u_{R,\,i=1,2,3}$ &  $\mathbf{\bar{3}}$  & $\mathbf{1}$ & $-2/3$ & $0$ \\
			\rule{0pt}{2.5ex} $d_{R,\,i=1,2,3}$ &  $\mathbf{\bar{3}}$  & $\mathbf{1}$ & $+1/3$ & $0$ \\
			\rule{0pt}{2.5ex} $h_u$ &  $\mathbf{1}$  & $\mathbf{2}$ & $+1/2$ & $0$ \\
			\rule{0pt}{2.5ex} $h_d$ &  $\mathbf{1}$  & $\mathbf{2}$ & $-1/2$ & $0$ \\
			\hline
			\rule{0pt}{2.5ex} $Q$ &  $\mathbf{3}$  & $\mathbf{2}$ & $+1/6$ & $-1$ \\
			\rule{0pt}{2.5ex} $Q'$ &  $\mathbf{\bar{3}}$  & $\mathbf{2}$ & $-1/6$ & $+1$ \\
			\rule{0pt}{2.5ex} $L$ &  $\mathbf{1}$  & $\mathbf{2}$ & $-1/2$ & $0$ \\
			\rule{0pt}{2.5ex} $L'$ &  $\mathbf{1}$  & $\mathbf{2}$ & $+1/2$ & $0$ \\
			\rule{0pt}{2.5ex} $E$ &  $\mathbf{1}$  & $\mathbf{1}$ & $+1$ & $0$ \\
			\rule{0pt}{2.5ex} $E'$ &  $\mathbf{1}$  & $\mathbf{1}$ & $-1$ & $0$ \\
			\rule{0pt}{2.5ex} $S_1$ &  $\mathbf{1}$  & $\mathbf{1}$ & $0$ & $-1$ \\
			\rule{0pt}{2.5ex} $S_2$ &  $\mathbf{1}$  & $\mathbf{1}$ & $0$ & $+1$ \\
			\hline
			\hline
		\end{tabular}
		\caption{The left-chiral superfields of the model and their gauge quantum numbers.}
		\label{tab:qn} 
	\end{center}
\end{table}

\bigskip
\noindent \textbf{SUSY sector masses}

The $10\times 10$ dimensional slepton mass matrix $M_{\tilde{e}}^2$, in the basis $(\tilde{e}_L,\tilde{e}_R,\tilde{\mu}_L,\tilde{\mu}_R, \tilde{\tau}_L,\tilde{\tau}_R,\tilde{E}_L, \tilde{E}_R, \tilde{E}', \tilde{E})$, reads: 
\be
M_{\tilde{e}_L\tilde{e}_L}^2=Y_{e,11}^2 v_d^2+m_{\tilde{e}_L}^2+m_Z^2\cos 2\beta\left(-\frac{1}{2}+\sin^2\theta_W\right)
\ee
\vspace{-0.4cm}
\be
M_{\tilde{e}_L\tilde{e}_R}^2=Y_{e,11} v_d \left(A_e-\mu\tanb\right)
\ee
\vspace{-0.4cm}
\be
M_{\tilde{e}_L\tilde{\mu}_L}^2=M_{\tilde{e}_L\tilde{\mu}_R}^2=M_{\tilde{e}_L\tilde{\tau}_L}^2=M_{\tilde{e}_L\tilde{\tau}_R}^2=M_{\tilde{e}_L\tilde{E}_L}^2=M_{\tilde{e}_L\tilde{E}_R}^2=M_{\tilde{e}_L\tilde{E}'}^2
=M_{\tilde{e}_L\tilde{E}}^2=0;
\ee
\smallskip
\vspace{-0.4cm}
\be
M_{\tilde{e}_R\tilde{e}_R}^2=Y_{e,11}^2 v_d^2+m_{\tilde{e}_L}^2-m_Z^2\cos 2\beta\,\sin^2\theta_W
\ee
\vspace{-0.4cm}
\be
M_{\tilde{e}_R\tilde{\mu}_L}^2=M_{\tilde{e}_R\tilde{\mu}_R}^2=M_{\tilde{e}_R\tilde{\tau}_L}^2=M_{\tilde{e}_R\tilde{\tau}_R}^2=M_{\tilde{e}_R\tilde{E}_L}^2=M_{\tilde{e}_R\tilde{E}_R}^2=M_{\tilde{e}_R\tilde{E}'}^2
=M_{\tilde{e}_R\tilde{E}}^2=0;
\ee
\smallskip
\vspace{-0.4cm}
\be
M_{\tilde{\mu}_L\tilde{\mu}_L}^2=Y_{e,22}^2v_d^2+m_{\tilde{\mu}_L}^2+\lam_{L,2}^2v_1^2+m_Z^2\cos 2\beta\left(-\frac{1}{2}+\sin^2\theta_W\right)-\frac{g_X^2}{2}\left(v_1^2-v_2^2\right)
\ee
\vspace{-0.4cm}
\be
M_{\tilde{\mu}_L\tilde{\mu}_R}^2=Y_{e,22}v_d\left(A_{\mu}-\mu\tanb\right)
\ee
\vspace{-0.4cm}
\be
M_{\tilde{\mu}_L\tilde{\tau}_L}^2=M_{\tilde{\mu}_L\tilde{\tau}_R}^2=0
\ee
\vspace{-0.4cm}
\be
M_{\tilde{\mu}_L\tilde{E}_L}^2=-M_L\,\lam_{L,2}v_1
\ee
\vspace{-0.4cm}
\be
M_{\tilde{\mu}_L\tilde{E}_R}^2=a_{S_1}^{(2)}v_1+\mu_S \lam_{L,2} v_2
\ee
\vspace{-0.4cm}
\be
M_{\tilde{\mu}_L\tilde{E}'}^2=-Y_{e,22}v_d\,\lam_{E,2} v_2+\widetilde{Y}_2 v_u \lam_{L,2} v_1
\ee
\vspace{-0.4cm}
\be
M_{\tilde{\mu}_L\tilde{E}}^2=0;
\ee
\smallskip
\vspace{-0.4cm}
\be
M_{\tilde{\mu}_R\tilde{\mu}_R}^2=Y_{e,22}^2 v_d^2+m_{\tilde{\mu}_R}^2+\lam_{E,2}^2 v_2^2-m_Z^2\cos 2\beta \sin^2\theta_W+\frac{g_X^2}{2}\left(v_1^2-v_2^2\right)
\ee
\vspace{-0.4cm}
\be
M_{\tilde{\mu}_R\tilde{\tau}_L}^2=M_{\tilde{\mu}_R\tilde{\tau}_R}^2=0
\ee
\vspace{-0.4cm}
\be
M_{\tilde{\mu}_R\tilde{E}_L}^2=0
\ee
\vspace{-0.4cm}
\be
M_{\tilde{\mu}_R\tilde{E}_R}^2=Y_{e,22}v_d\,\lam_{L,2} v_1-\widetilde{Y}_2 v_u \lam_{E,2} v_2
\ee
\vspace{-0.4cm}
\be
M_{\tilde{\mu}_R\tilde{E}'}^2=-a_{S_2}^{(2)}v_2-\mu_S \lam_{E,2} v_1
\ee
\vspace{-0.4cm}
\be
M_{\tilde{\mu}_R\tilde{E}}^2=-M_E\,\lam_{E,2}v_2;
\ee
\smallskip
\vspace{-0.4cm}
\be
M_{\tilde{\tau}_L\tilde{\tau}_L}^2=Y_{e,33}^2 v_d^2+m_{\tilde{\tau}_L}^2+m_Z^2\cos 2\beta\left(-\frac{1}{2}+\sin^2\theta_W\right)+\frac{g_X^2}{2}\left(v_1^2-v_2^2\right)
\ee
\vspace{-0.4cm}
\be
M_{\tilde{\tau}_L\tilde{\tau}_R}^2=Y_{e,33} v_d \left(A_{\tau}-\mu\tanb\right)
\ee
\vspace{-0.4cm}
\be
M_{\tilde{\tau}_L\tilde{E}_L}^2=M_{\tilde{\tau}_L\tilde{E}_R}^2=M_{\tilde{\tau}_L\tilde{E}'}^2
=M_{\tilde{\tau}_L\tilde{E}}^2=0;
\ee
\smallskip
\vspace{-0.4cm}
\be
M_{\tilde{\tau}_R\tilde{\tau}_R}^2=Y_{e,33}^2 v_d^2+m_{\tilde{\tau}_L}^2-m_Z^2\cos 2\beta\,\sin^2\theta_W-\frac{g_X^2}{2}\left(v_1^2-v_2^2\right)
\ee
\vspace{-0.4cm}
\be
M_{\tilde{\tau}_R\tilde{E}_L}^2=M_{\tilde{\tau}_R\tilde{E}_R}^2=M_{\tilde{\tau}_R\tilde{E}'}^2
=M_{\tilde{\tau}_R\tilde{E}}^2=0;
\ee
\smallskip
\vspace{-0.4cm}
\be
M_{\tilde{E}_L\tilde{E}_L}^2=m_L^2+M_L^2+\widetilde{Y}_1^2 v_d^2+m_Z^2\cos 2\beta\left(-\frac{1}{2}+\sin^2\theta_W\right)
\ee
\vspace{-0.4cm}
\be
M_{\tilde{E}_L\tilde{E}_R}^2=-B_{M_L}
\ee
\vspace{-0.4cm}
\be
M_{\tilde{E}_L\tilde{E}'}^2=-M_L\widetilde{Y}_2 v_u +M_E\widetilde{Y}_1 v_d
\ee
\vspace{-0.4cm}
\be
M_{\tilde{E}_L\tilde{E}}^2=\widetilde{Y}_1 v_d\left(A_{Y_1}-\mu\tanb\right);
\ee
\smallskip
\vspace{-0.4cm}
\be
M_{\tilde{E}_R\tilde{E}_R}^2=m_{L'}^2+M_L^2+\lam_{L,2}^2 v_1^2+\widetilde{Y}_2^2 v_u^2-m_Z^2\cos 2\beta\left(-\frac{1}{2}+\sin^2\theta_W\right)
\ee
\vspace{-0.4cm}
\be
M_{\tilde{E}_R\tilde{E}'}^2=\widetilde{Y}_2 v_u\left(A_{Y_2}-\mu\cot\beta\right)
\ee
\vspace{-0.4cm}
\be
M_{\tilde{E}_R\tilde{E}}^2=M_E\widetilde{Y}_2 v_u-M_L\widetilde{Y}_1 v_d\,;
\ee
\smallskip
\vspace{-0.4cm}
\be
M_{\tilde{E}'\tilde{E}'}^2=m_{E'}^2+M_E^2+\lam_{E,2}^2 v_2^2+\widetilde{Y}_2^2 v_u^2+m_Z^2\cos 2\beta \sin^2\theta_W
\ee
\vspace{-0.4cm}
\be
M_{\tilde{E}'\tilde{E}}^2= B_{M_E}\,;
\ee
\smallskip
\vspace{-0.4cm}
\be
M_{\tilde{E}\tilde{E}}^2=m_E^2+M_E^2+\widetilde{Y}_1^2 v_d^2-m_Z^2\cos 2\beta \sin^2\theta_W.
\ee
\smallskip

The slepton mass matrix is diagonalized by a unitary matrix $R^s$, such that 
\be
\mathcal{D}_{\tilde{e}}^2=R^s M_{\tilde{e}}^2R^{s \dag}
\ee 
and the mass eigenstates are given in terms of the gauge eigenstates, in the order given above, as 
\be
\tilde{e}_i=R_{ij}^s\,\tilde{e}_j'\,.
\ee

And finally let us explicitly write the $7\times 7$ dimensional 
neutralino mass matrix at the tree level, in the basis $(\tilde{B},\tilde{W},\tilde{h}_d,\tilde{h}_u,\tilde{Z}',\tilde{S}_1,\tilde{S}_2)$,
\be
M_{\chi}=\left(\begin{array}{ccccccc}
	M_1 & 0 & -\frac{1}{\sqrt{2}} g_1 v_d & \frac{1}{\sqrt{2}} g_1 v_u & 0 & 0 & 0  \\
	0 & M_2 & \frac{1}{\sqrt{2}} g_2 v_d & -\frac{1}{\sqrt{2}} g_2 v_u & 0 & 0 & 0  \\
	-\frac{1}{\sqrt{2}} g_1 v_d & \frac{1}{\sqrt{2}} g_2 v_d & 0 & -\mu & 0 & 0 & 0  \\
	\frac{1}{\sqrt{2}} g_1 v_u & \frac{1}{\sqrt{2}} g_2 v_u & -\mu & 0 & 0 & 0 & 0  \\
	0 & 0 & 0 & 0 & M_{\tilde{Z}'} & -g_X v_1 & g_X v_2  \\
	0 & 0 & 0 & 0 & -g_X v_1 & 0 & \mu_S  \\
	0 & 0 & 0 & 0 & g_X v_2 & \mu_S & 0 \\
\end{array}\right).
\ee
The neutralino mass matrix is diagonalized by a unitary matrix $R^n$, such that 
\be
\mathcal{D_{\chi}}=R^{n \ast}M_{\chi}R^{n \dag}
\ee 
and the mass eigenstates are given in terms of the gauge eigenstates, in the order given above, as 
\be
\chi_i^0=R_{ij}^n\,\chi_j'\,.
\ee

\bigskip
\noindent \textbf{Higgs sector masses}

We neglect in this work the small kinetic mixing between the SU(2)$_L\times$U(1)$_Y$ and U(1)$_X$ gauge bosons. 
As a consequence, the scalar sector associated with the breaking of U(1)$_X$ mixes with the MSSM Higgs sector at the loop level only. At the tree level, the squared 
mass matrices of the Higgs scalars are thus block-diagonal and the physical fields can be easily parameterized in terms of the gauge eigenstates as
\be
\left( {\begin{array}{c}
		H\\
		h
\end{array} } \right) =\left( {\begin{array}{cc}
		\cos\alpha & \sin\alpha\\
		-\sin\alpha & \cos\alpha
\end{array} } \right) \left( {\begin{array}{c}
		\Re[h_d]\\
		\Re[h_u]
\end{array} } \right), \quad \left( {\begin{array}{c}
		H'\\
		h'
\end{array} } \right) =\left( {\begin{array}{cc}
		\cos\alpha' & \sin\alpha'\\
		-\sin\alpha' & \cos\alpha'
\end{array} } \right) \left( {\begin{array}{c}
		\Re[S_1]\\
		\Re[S_2]
\end{array} } \right),
\ee 
where the angles $\alpha$, $\alpha'$, which explicitly enter Eqs.~(\ref{clscal})-(\ref{crscal}), parameterize, respectively, rotation matrices $R_{ij}^{h\ast}$ and $R_{ij}^{h'\ast}$ that diagonalize the scalar mass matrices squared of the two sectors:
\be
\mathcal{D}_h^2=R^h M_{h,2\times 2}^2 R^{h\dag}, \quad \mathcal{D}_{h'}^2=R^{h'} M_{h',2\times 2}^2 R^{h'\dag}.\label{scalmat}
\ee
Note that the explicit form of the Higgs mass matrix, $M_{h,2\times 2}^2$ in \refeq{scalmat}, is the same as in the MSSM, whereas $M_{h',2\times 2}^2$ can be constructed in a specular way and, not being particularly illuminating,  we 
refrain from presenting it explicitly here.

Equivalently, we parameterize the new pseudoscalar field, $A'$, and Goldstone boson $G'^0$ as
\be
\left( {\begin{array}{c}
		A'\\
		G'^{0}
\end{array} } \right) =\left( {\begin{array}{cc}
		\cos\alpha'' & \sin\alpha''\\
		-\sin\alpha'' & \cos\alpha''
\end{array} } \right) \left( {\begin{array}{c}
		i\,\Im[S_1]\\
		i\,\Im[S_2]
\end{array} } \right),
\ee 
in terms of the effective angle $\alpha''$ which is also featured in Eqs.~(\ref{clscal})-(\ref{crscal}).

\bigskip
\noindent \textbf{Tadpole equations}

Finally, we recall the tree-level tadpole equations resulting from minimization of the scalar potential with respect to the extra sector scalar fields $S_1$ and $S_2$:
\bea
B_{\mu_S}&=&\frac{\left(-2\mu_S^2-m^2_{S_1}-m^2_{S_2}\right)\sin 2\beta_S}{2}\,,\nonumber\\
\mu_S^2&=&-\frac{m^2_{Z'}}{2}-\frac{\tan^2\beta_S\,m^2_{S_1}-m^2_{S_2}}{\tan^2\beta_S-1}\,,
\eea
in terms of the superpotential and soft SUSY-breaking parameters defined in Eqs.~(\ref{superpot}) and~(\ref{softlagr}), and $\tanb_S=v_1/v_2$.

\bigskip
\bibliographystyle{JHEP}
\bibliography{mybib}

\end{document}

%% file: macros_SUS.tex
\newcommand{\newc}{\newcommand*}

\long\def\begincomment#1\endcomment{%
        \begingroup\sf\baselineskip12pt#1\endgroup}

\newc{\etal}{\textrm{et al.}} 
\newc{\eg}{\textrm{e.g.}} 
\newc{\ie}{\textrm{i.e.}}
\newc{\etc}{\textrm{etc}}
\newc\vs{\textrm{vs.}}
\newc{\cl}{\rm {C.L.}}

\newc{\ev}{\ensuremath{\,\mathrm{eV}}}
\newc{\kev}{\ensuremath{\,\mathrm{keV}}}
\newc{\mev}{\ensuremath{\,\mathrm{MeV}}}
\newc{\gev}{\ensuremath{\,\mathrm{GeV}}}
\newc{\tev}{\ensuremath{\,\mathrm{TeV}}}
\newc{\MeV}{\mev} 
\newc{\TeV}{\tev}
\newc{\invpb}{\ensuremath{/\text{pb}}}
\newc{\invfb}{\ensuremath{\,\textrm{fb}^{-1}}}
\newc\nb{\ensuremath{\,\mathrm{nb}}} \newc\pb{\ensuremath{\,\mathrm{pb}}} \newc\fb{\ensuremath{\,\mathrm{fb}}}
\newc\pc{\ensuremath{\,\mathrm{pc}}}
\newc\kpc{\ensuremath{\,\mathrm{kpc}}}
\newc\mpc{\ensuremath{\,\mathrm{Mpc}}}
\newc\ps{\ensuremath{\,\mathrm{ps}}} 
\newc\cmeter{\ensuremath{\,\mathrm{cm}}} 
\newc\meter{\ensuremath{\,\mathrm{m}}} 
\newc\kmeter{\ensuremath{\,\mathrm{km}}}
\newc\second{\ensuremath{\,\mathrm{s}}}
\newc\msecond{\ensuremath{\,\mathrm{ms}}}
\newc\nsecond{\ensuremath{\,\mathrm{ns}}}
\newc\psecond{\ensuremath{\,\mathrm{ps}}}

\newc{\chisqmin}{\ensuremath{\chi^2_{\mathrm{min}}}}
\newc{\Delchisq}{\ensuremath{\Delta\chi^2}}
\newc{\chisq}{\ensuremath{\chi^2}}
\newc{\like}{\ensuremath{\mathcal{L}}}

\newc\lsim{\ensuremath{\mathrel{\rlap{\lower4pt\hbox{\hskip1pt$\sim$}}\raise1pt\hbox{$<$}}}}
\newc\gsim{\ensuremath{\mathrel{\rlap{\lower4pt\hbox{\hskip1pt$\sim$}}\raise1pt\hbox{$>$}}}}
\newc{\VEV}[1]{\ensuremath{\langle #1 \rangle}}
\newc{\dl}{\ensuremath{\stackrel{\leftarrow}{D}}}
\newc{\dr}{\ensuremath{\stackrel{\rightarrow}{D}}}

\newc{\bcenter}{\begin{center}}    \newc{\ecenter}{\end{center}}
\newc{\bfl}{\begin{flushleft}}    \newc{\efl}{\end{flushleft}}
\newc{\bfr}{\begin{flushright}}    \newc{\efr}{\end{flushright}}

\newc{\bi}{\begin{itemize}}
\newc{\ei}{\end{itemize}}
\newc{\bed}{\begin{description}}
\newc{\eed}{\end{description}}
\newc{\ben}{\begin{enumerate}}
\newc{\een}{\end{enumerate}}

\newc{\be}{\begin{equation}}
\newc{\ee}{\end{equation}}
\newc{\bea}{\begin{eqnarray}}
\newc{\eea}{\end{eqnarray}}
\newc{\bfle}{\begin{flalign}}
\newc{\efle}{\end{flalign}}
\newc{\ra}{\rightarrow}

\newc{\alphas}{\ensuremath{\alpha_s}}
\newc{\alphatwo}{\ensuremath{\alpha_2}}
\newc{\alphaone}{\ensuremath{\alpha_1}}
\newc{\alphai}[1]{\ensuremath{\alpha_{#1}}}
\newc{\alphaem}{\ensuremath{\alpha_{\mathrm{em}}}}
\newc{\alphaeff}{\ensuremath{\alpha_{\mathrm{eff}}}}
\newc{\sineff}{\ensuremath{\sin \theta_{\mathrm{eff}}}}
\newc{\sinsqeff}{\ensuremath{\sin^2 \theta_{\mathrm{eff}}}}
\newc{\dalphahad}{\ensuremath{\Delta \alpha_{\mathrm{had}}}}
\newc{\yt}{\ensuremath{h_t}} \newc{\yb}{\ensuremath{h_b}} \newc{\ytau}{\ensuremath{h_{\tau}}}
\newc\mz{\ensuremath{M_Z}} 
\newc\mw{\ensuremath{m_W}}
\newc\mZ{\mz}        \newc\mW{\mw}
\newc\mhsm{\ensuremath{ m_{H_{\mathrm{SM}}}}}
\newc{\mtop}{\ensuremath{ m_t}}               \newc{\mtpole}{\ensuremath{ M_t}}
\newc{\mbottom}{\ensuremath{ m_b}} 
\newc{\mtau}{\ensuremath{ m_{\tau}}}
\newc{\mt}{\mtpole}
\newc{\mb}{\mbottom} 
\newc{\rtwogg}{\ensuremath{R_{h_2}(\gamma\gamma)}}
\newc{\rtwozz}{\ensuremath{R_{h_2}(ZZ)}}
\newc{\ronegg}{\ensuremath{R_{h_1}(\gamma\gamma)}}
\newc{\ronezz}{\ensuremath{R_{h_1}(ZZ)}}
\newc{\rsiggg}{\ensuremath{R_{h_\textrm{sig}}(\gamma\gamma)}}
\newc{\rsigzz}{\ensuremath{R_{h_\textrm{sig}}(ZZ)}}
\newc{\llbar}{\ensuremath{\ell\bar{\ell}}}
\newc{\tauptaum}{\ensuremath{ \tau^+\tau^-}}
\newc{\qqbar}{\ensuremath{ q\bar{q}}} \newc{\ppbar}{\ensuremath{ p\bar{p}}}
\newc{\bbbar}{\ensuremath{ b\bar{b}}} \newc{\ttbar}{\ensuremath{ t\bar{t}}}
\newc{\ffbar}{\ensuremath{ f\bar{f}}} \newc{\tautaubar}{\ensuremath{ \tau\bar{\tau}}}

\newc{\mchi}{\ensuremath{m_\neutone}}
\newc{\squark}{\ensuremath{\tilde{q}}}
\newc{\slepton}{\ensuremath{\tilde{l}}}
\newc{\gluino}{\ensuremath{\tilde{g}}} 
\newc{\mgluino}{\ensuremath{{m_{\gluino}}}}
\newc{\wino}{\ensuremath{\tilde{W}}} 
\newc{\mwino}{\ensuremath{{m_{\wino}}}}
\newc{\tone}{\ensuremath{{\tilde{t}_1}}}
\newc{\bone}{\ensuremath{{\tilde{b}_1}}}
\newc{\Hone}{\ensuremath{{\tilde{H}_{1}}}}
\newc{\Htwo}{\ensuremath{{\tilde{H}_{2}}}}
\newc{\Hhtwo}{\ensuremath{{H_{2}}}}
\newc{\qli}{\ensuremath{{\tilde{Q}_{i}}}}
\newc{\uri}{\ensuremath{{\tilde{u}_{i}}}}
\newc{\dri}{\ensuremath{{\tilde{d}_{i}}}}
\newc{\lli}{\ensuremath{{\tilde{L}_{i}}}}
\newc{\eri}{\ensuremath{{\tilde{e}_{i}}}}

\newc{\sthw}{\ensuremath{ \sin\theta_W}}              \newc{\cthw}{\ensuremath{\cos\theta_W}}
\newc{\tanthw}{\ensuremath{ \tan\theta_W}}              \newc{\cotthw}{\ensuremath{\cot\theta_W}}
\newc{\ssqthw}{\ensuremath{\sin^2 \theta_W}}
\newc{\msbar}{\ensuremath{\overline{MS}}} \newc{\drbar}{\ensuremath{\overline{DR}}}
\newc{\mtmtsmmsbar}{\ensuremath{ m_t(m_t)^{\msbar}_{{\mathrm{SM}}}}}
\newc{\mtmtsmdrbar}{\ensuremath{ m_t(m_t)^{\drbar}_{{\mathrm{SM}}}}}
\newc{\mtmtmssmdrbar}{\ensuremath{ m_t(m_t)^{\drbar}_{{\mathrm{SUSY}}}}}
\newc{\mbmbmsbar}{\ensuremath{ m_b(m_b)^{\msbar} }}
\newc{\mbmbsmmsbar}{\ensuremath{ m_b(m_b)^{\msbar}_{{\mathrm{SM}}}}}
\newc{\mbmzsmmsbar}{\ensuremath{ m_b(\mz)^{\msbar}_{{\mathrm{SM}}}}}
\newc{\mbmzsmdrbar}{\ensuremath{ m_b(\mz)^{\drbar}_{{\mathrm{SM}}}}}
\newc{\mbmzmssmdrbar}{\ensuremath{ m_b(\mz)^{\drbar}_{{\mathrm{SUSY}}}}}
\newc{\mtaumzsmmsbar}{\ensuremath{ m_{\tau}(\mz)^{\msbar}_{{\mathrm{SM}}}}}
\newc{\mtaumzsmdrbar}{\ensuremath{ m_{\tau}(\mz)^{\drbar}_{{\mathrm{SM}}}}}
\newc{\mtaumzmssmdrbar}{\ensuremath{ m_{\tau}(\mz)^{\drbar}_{{\mathrm{SUSY}}}}}
\newc{\alphasmzms}{\ensuremath{\alpha_s(M_Z)^{\overline{MS}}}}
\newc{\alphaimzms}[1]{\ensuremath{\alpha_{#1}(M_Z)^{\overline{MS}}}}
\newc{\alphaemmz}{\ensuremath{\alpha_{\mathrm{em}}(M_Z)^{\overline{MS}}}}

\newc{\mzero}{\ensuremath{{m_0}}}
\newc{\mhalf}{\ensuremath{ m_{1/2}}}
\newc{\tanb}{\ensuremath{\tan\beta}}
\newc{\azero}{\ensuremath{ A_0}}
\newc{\signmu}{\ensuremath{\rm{sgn}\,\mu}}
\newc{\atau}{\ensuremath{{A_{\tau}}}}
\newc{\mueff}{\ensuremath{\mu_{\rm{eff}}}}
\newc{\lam}{\ensuremath{{\lambda}}}
\newc{\kap}{\ensuremath{{\kappa}}}
\newc{\alam}{\ensuremath{{A_{\lambda}}}}
\newc{\akap}{\ensuremath{{A_{\kappa}}}}
\newc{\hs}{\ensuremath{ H_s}}      
\newc{\mhs}{\ensuremath{ m_{H_s}}} 
\newc{\mgut}{\ensuremath{ M_{\rm GUT}}}
\newc{\gut}{\ensuremath{{\rm GUT}}}
\newc{\mplanck}{\ensuremath{ M_{\rm P}}}      \newc{\mpl}{\ensuremath{ M_{\rm Pl}}}
\newc{\msusy}{\ensuremath{ M_{\rm SUSY}}}      \newc{\ms}{\ensuremath{ M_{\rm S}}}
 \newc{\hu}{\ensuremath{ H_u}}       \newc{\hd}{\ensuremath{ H_d}}
 \newc{\mhu}{\ensuremath{ m_{H_u}}}       \newc{\mhd}{\ensuremath{ m_{H_d}}}
 \newc{\mhuew}{\ensuremath{ m^{\ast}_{H_u}}}       \newc{\mhdew}{\ensuremath{ m^{\ast}_{H_d}}}
 \newc{\mhuewsq}{\ensuremath{ m^{\ast\, 2}_{H_u}}}       \newc{\mhdewsq}{\ensuremath{ m^{\ast\, 2}_{H_d}}}
 \newc{\mhl}{\ensuremath{m_\hl}} 
 \newc{\mhone}{\ensuremath{m_{h_1}}} 
 \newc{\mhtwo}{\ensuremath{m_{h_2}}} 
 \newc{\mhi}{\ensuremath{m_{\tilde{h}}}} 
 \newc{\mul}{\ensuremath{m_{\tilde{u}_L}}} 
 \newc{\mbone}{\ensuremath{m_{\tilde{b}_1}}}  
 \newc{\mtone}{\ensuremath{m_{\tilde{t}_1}}} 
 \newc{\ma}{\ensuremath{m_A}} 
 \newc{\mH}{\ensuremath{m_H}} 
 \newc{\maone}{\ensuremath{m_{a_1}}} 
 \newc{\matwo}{\ensuremath{m_{a_2}}}
 \newc{\hone}{\ensuremath{h_1}}
 \newc{\htwo}{\ensuremath{h_2}}
 \newc{\aone}{\ensuremath{a_1}}
 \newc{\atwo}{\ensuremath{a_2}}
 \newc{\mqthree}{\ensuremath{m_{\tilde{Q}_3}^2}}
 \newc{\muthree}{\ensuremath{m_{\tilde{u}_3}^2}}
 \newc{\mqli}{\ensuremath{m_{\tilde{Q}_{i}}}}
 \newc{\muri}{\ensuremath{m_{\tilde{u}_{i}}}}
 \newc{\mdri}{\ensuremath{m_{\tilde{d}_{i}}}}
 \newc{\mlli}{\ensuremath{m_{\tilde{L}_{i}}}}
 \newc{\meri}{\ensuremath{m_{\tilde{e}_{i}}}}
 \newc{\ts}{\ensuremath{T_{SUSY}}}

\newc{\sigsip}{\ensuremath{\sigma^{\rm SI}_{p}}}	\newc{\sigsin}{\ensuremath{\sigma^{\rm SI}_{n}}}
\newc{\sigsdp}{\ensuremath{\sigma^{\rm SD}_{p}}}	\newc{\sigsdn}{\ensuremath{\sigma^{\rm SD}_{n}}}
\newc{\sigsi}{\ensuremath{\sigma^{\rm SI}}}	\newc{\sigsd}{\ensuremath{\sigma^{\rm SD}}}
\newc{\abund}{\ensuremath{ \Omega h^2}}
\newc{\omegadm}{\ensuremath{ \Omega_{{\rm DM}}}}     \newc{\abunddm}{\ensuremath{ \Omega_{{\rm DM}} h^2}} 
\newc{\omegam}{\ensuremath{ \Omega_{{\rm m}}}}       \newc{\abundm}{\ensuremath{ \Omega_{{\rm m}} h^2}}
\newc{\omegab}{\ensuremath{ \Omega_{{\rm b}}}}	\newc{\abundb}{\ensuremath{ \Omega_{{\rm b}} h^2}}
\newc{\omegatot}{\ensuremath{ \Omega_{{\rm TOT}}}}
\newc{\omegacdm}{\ensuremath{ \Omega_{{\rm CDM}}}}   \newc{\abundcdm}{\ensuremath{ \Omega_{{\rm CDM}} h^2}}
\newc{\omegalambda}{\ensuremath{ \Omega_{\Lambda}}} \newc{\abundlambda}{\ensuremath{ \Omega_{\Lambda} h^2}}
\newc{\omegarad}{\ensuremath{ \Omega_{{\rm rad}}}}  \newc{\abundrad}{\ensuremath{ \Omega_{{\rm rad}} h^2}}
\newc{\rhocrit}{\ensuremath{ \rho_{\rm crit}}}
\newc{\rhochi}{\ensuremath{ \rho_{\chi}}}
\newc{\abunchi}{\ensuremath{\Omega_\chi h^2}}
\newc{\abundlsp}{\ensuremath{\Omega_{\rm LSP}h^2}}
\newcommand*{\abundchi}{\ensuremath{\Omega_\chi h^2}}

\newc{\amu}{\ensuremath{ a_{\mu}}}        \newc{\amususy}{\ensuremath{ a_{\mu}^{\mathrm{SUSY}}}}
\newc{\amuexpt}{\ensuremath{ a_{\mu}^{\mathrm{expt}}}}        \newc{\amusm}{\ensuremath{ a_{\mu}^{\mathrm{SM}}}}
\newc\deltaamu{\ensuremath{\Delta a_{\mu}}} \newc{\deltaamususy}{\ensuremath{\delta a_{\mu}^{\mathrm{SUSY}}}}
\newc\gmtwo{\ensuremath{ (g-2)_{\mu}}} 
\newc{\deltagmtwomususy}{\ensuremath{\delta\left(g-2\right)_{\mu}^{\mathrm{SUSY}}}}
\newc{\deltagmtwomu}{\ensuremath{\delta\left(g-2\right)_{\mu}}}
\newc\BR{\ensuremath{\rm BR}}
\newc\bsgamma{\ensuremath{ b\rightarrow s \gamma }}
\newc\bxsgamma{\ensuremath{\overline{B}\rightarrow X_{s}\gamma}}
\newc\brbsgamma{\ensuremath{\BR\left(\bsgamma\right)}}
\newc\brbxsgamma{\ensuremath{\BR\left(\bxsgamma\right)}}
\newc\bsmumu{\ensuremath{B_s\to\mu^+\mu^-}}
\newc\brbsmumu{\ensuremath{\BR\left(B_s\to\mu^+\mu^-\right)}}
\newc\bdmmumu{\ensuremath{\overline{B}_d\to\mu^+\mu^-}}
\newc\bbbarmix{\ensuremath{\overline{B}_s\mbox{-}B_s}}      
\newc\delmbs{\ensuremath{\Delta M_{B_s}}}
\newc{\butaunu}{\ensuremath{B_u \rightarrow \tau \nu}}
\newc{\brbutaunu}{\ensuremath{\BR\left(B_u \rightarrow \tau \nu\right)}}

\newcommand*{\reftable}[1]{Table~\ref{#1}}         \newcommand*{\reftables}[2]{Tables~\ref{#1} and \ref{#2} }
       
\newcommand*{\reffig}[1]{Fig.~\ref{#1}}
 
        \newcommand*{\refeq}[1]{Eq.~(\ref{#1})}
     \newcommand*{\refsec}[1]{Sec.~\ref{#1}}

\newcommand*{\neutone}{\ensuremath{\chi^0_1}}
\newcommand*{\neuttwo}{\ensuremath{\tilde{{\chi}}^0_2}}

\newcommand*{\charone}{\ensuremath{{\chi}^{\pm}_1}}

\newcommand*{\madgr}{\texttt{MadGraph5\_aMC@NLO}}

\newcommand*{\chep}{{\tt CalcHEP}}
\newcommand*{\spheno}{{\tt SPheno}}

\let\oldcite\cite
\renewcommand*{\cite}{~\oldcite}

\newcommand*{\hl}{\ensuremath{h}}

\newcommand*{\rk}{\ensuremath{R_K}}
\newcommand*{\rks}{\ensuremath{R_{K^{\ast}}}}
\newcommand*{\zp}{\ensuremath{Z'}}

\newcommand*{\hp}{\ensuremath{h'}}